\documentclass[journal=jcp,manuscript=article]{achemso}
\setkeys{acs}{maxauthors=0}

\usepackage{achemso}
\usepackage{graphics}
\usepackage{amssymb,amsfonts}
\usepackage{graphicx}
\usepackage[table]{xcolor}
\usepackage{caption}
\usepackage{subcaption}
\usepackage{colortbl}
\usepackage{amsmath}
\usepackage{amsopn}
\usepackage{bm}
\usepackage{color}
\usepackage{array}
\usepackage{lscape}
\usepackage{mciteplus}
\usepackage[version=3]{mhchem}
\usepackage{ulem}
\usepackage{listings}
\usepackage{enumerate}
\SectionNumbersOn
\makeatletter
\newcommand*\rel@kern[1]{\kern#1\dimexpr\macc@kerna}
\newcommand*\widebar[1]{%
  \begingroup
  \def\mathaccent##1##2{%
    \rel@kern{0.8}%
    \overline{\rel@kern{-0.8}\macc@nucleus\rel@kern{0.2}}%
    \rel@kern{-0.2}%
  }%
  \macc@depth\@ne
  \let\math@bgroup\@empty \let\math@egroup\macc@set@skewchar
  \mathsurround\z@ \frozen@everymath{\mathgroup\macc@group\relax}%
  \macc@set@skewchar\relax
  \let\mathaccentV\macc@nested@a
  \macc@nested@a\relax111{#1}%
  \endgroup
}
\makeatother
\numberwithin{equation}{subsection}

\author{Janus J. Eriksen}
\email{janusje@chem.au.dk}
\affiliation[Aarhus University]
{qLEAP Center for Theoretical Chemistry, Department of Chemistry, Aarhus University, DK-8000 Aarhus C, Denmark}
\author{Devin A. Matthews}
\affiliation[The University of Texas at Austin]
{The Institute for Computational Engineering and Sciences, The University of Texas at Austin, Austin, Texas 78712, United States}
\author{Poul J\o rgensen}
\affiliation[Aarhus University]
{qLEAP Center for Theoretical Chemistry, Department of Chemistry, Aarhus University, DK-8000 Aarhus C, Denmark}
\author{J\"urgen Gauss}
\affiliation[Johannes Gutenberg-Universit\"at Mainz]
{Institut f\"ur Physikalische Chemie, Johannes Gutenberg-Universit\"at Mainz, D-55128 Mainz, Germany}

\title[TITLE]
  {Assessment of the accuracy of coupled cluster perturbation theory for open-shell systems. I. Triples expansions}

\begin{document}
%
%
\begin{abstract}

The accuracy at which total energies of open-shell atoms and organic radicals may be calculated is assessed for selected coupled cluster perturbative triples expansions, all of which augment the coupled cluster singles and doubles (CCSD) energy by a non-iterative correction for the effect of triple excitations. Namely, the second- through sixth-order models of the recently proposed CCSD(T--$n$) triples series [J. Chem. Phys. {\bf{140}}, 064108 (2014)] are compared to the acclaimed CCSD(T) model for both unrestricted as well as restricted open-shell Hartree-Fock (UHF/ROHF) reference determinants. By comparing UHF- and ROHF-based statistical results for a test set of 18 modest-sized open-shell species with comparable RHF-based results, no behavioral differences are observed for the higher-order models of the CCSD(T--$n$) series in their correlated descriptions of closed- and open-shell species. In particular, we find that the convergence rate throughout the series towards the coupled cluster singles, doubles, and triples (CCSDT) solution is identical for the two cases. For the CCSD(T) model, on the other hand, not only its numerical consistency, but also its established, yet fortuitous cancellation of errors breaks down in the transition from closed- to open-shell systems. The higher-order CCSD(T--$n$) models (orders $n>3$) thus offer a consistent and significant improvement in accuracy relative to CCSDT over the CCSD(T) model, equally for RHF, UHF, and ROHF reference determinants, albeit at an increased computational cost.

%
\end{abstract}
\newpage
%

%
%
\section{Introduction}\label{intro_section}

The coupled cluster (CC) hierarchy of size-extensive correlated methods~\cite{cicek_1,*cicek_2,*paldus_cikek_shavitt,mest,shavitt_bartlett_cc_book} features some of the most prominent {\it{ab initio}} wave function-based models in use today, namely the CC singles and doubles (CCSD) model~\cite{ccsd_paper_1_jcp_1982}, the CC singles, doubles, and triples (CCSDT) model~\cite{ccsdt_paper_1_jcp_1987,*ccsdt_paper_2_cpl_1988}, the CC singles, doubles, triples, and quadruples (CCSDTQ) model~\cite{ccsdtq_paper_1_jcp_1991,*ccsdtq_paper_2_jcp_1992}, etc. In the limit of correlation between all $N$ electrons of a molecular system, the full configuration interaction (FCI) energy is returned. While the iterative CC models offer excellent precision, the practical usefulness---in particular of the higher-level models for modest- and large-sized systems---is hampered by a polynomial increase of the computational cost up through the hierarchy. For this reason, a related branch of correlated methods, which incorporate non-iterative many-body perturbation theory~\cite{shavitt_bartlett_cc_book} (MBPT) energy corrections into the CC treatment, has been devised. However, while these perturbative corrections lower the overall cost of the computation, a proper iterative description of the effects of single and double excitations (at times even triple and quadruple excitations) within a CC scheme is called for, which is why methods have been devised in which CC and perturbation theory are coupled. The most renowned of such combined methods is the CCSD(T) model~\cite{original_ccsdpt_paper}, in which the iterative CCSD energy is augmented by selected MBPT-rationalized corrections for triple excitations~\cite{ccsdpt_perturbation_stanton_cpl_1997}, and owing in part to fortuitous, yet remarkably consistent error cancellations and in part a favorable computational scaling, CCSD(T) has developed into the model of choice whenever high-accuracy total, reaction, and interaction energies, equilibrium geometries, and various other molecular properties are desired, cf., for instance, the calibration chapter of Ref. \citenum{mest}.

The authoritative position held today by the CCSD(T) model has been justified by the results of a great number of numerical studies and experimental comparisons, of which the clear majority has been based on closed-shell molecules in their equilibrium geometry, despite the fact that UHF- and ROHF-based CCSD(T) implementations have been available almost from the beginning. However, being able to accurately account for triples effects in closed-shell systems will only provide access to a fraction of the manifold of chemical relevant problems encountered in, e.g., modern catalysis, biological, or even material sciences. Within the diversified world of reaction-based chemistry, for instance, radicals and bare atoms are omnipresent. Since such reactive species are notoriously difficult to characterize experimentally, computational chemistry, preferably of sufficiently high accuracy, hence has a substantial and valuable role to play in their description and characterization. Thus, taking into account today's automatic predilection for the CCSD(T) model, it becomes vital to pose the following question: is it obvious that the striking cancellation of errors, which gives rise to the stability of the model, necessarily carries over from closed- to open-shell calculations? In answering this, it appears essential to rigorously assess to what accuracy CCSD(T) total energies of atoms and organic radicals can be obtained, and, if this level of accuracy is not qualitatively acceptable, establish new methods and protocols that enable a balanced treatment of both closed- and open-shell species. On this note, it has previously been shown how the precision of equilibrium geometries and harmonic vibrational frequencies of small- to modest-sized molecular radicals, as calculated by the CCSD(T) model, do not mirror that found for comparable closed-shell molecules~\cite{byrd_head_gordon_open_shell_cc_jpca_2001}. Furthermore, the notion of a CC/MBPT hierarchy of systematically improved models seems not to apply straightaway for open-shell molecules, as second-order M{\o}ller-Plesset (MP2)~\cite{mp2_phys_rev_1934} results are often no improvement over the underlying mean-field Hartree-Fock (HF) solution~\cite{byrd_head_gordon_open_shell_cc_jpca_2001,beran_head_gordon_open_shell_cc_pccp_2003}. 

In the present work, we wish to suggest and advocate---as an alternative to the widely-used CCSD(T) model---the recently proposed CCSD(T--$n$) models~\cite{ccsd_pert_theory_jcp_2014,eom_cc_pert_theory_jcp_2014,triples_pert_theory_jcp_2015}, which not only offer a convergent and well-mannered series of CC triples models for the accurate determination of total electronic energies, but do so, as we will demonstrate, regardless of the spin of the ground state. The CCSD(T--$n$) series is defined as an order expansion in the M{\o}ller-Plesset fluctuation potential of a bivariational CCSDT energy Lagrangian where, from a CCSD zeroth-order expansion point, perturbative solutions of both the exponential CCSDT cluster state and linear $\Lambda$-state are embedded into the energy corrections~\cite{e_ccsd_tn_jcp_2016}. In Ref. \citenum{triples_pert_theory_jcp_2015}, the accuracy of the second- through fourth-order models of the CCSD(T--$n$) series (the CCSD(T--2), CCSD(T--3), and CCSD(T--4) models, respectively) was statistically assessed for total electronic energies of a test set of 17 small closed-shell molecules through a comparison with alternative CC triples expansions, including the CCSD(T) model. Furthermore, non-parallelity errors with respect to the native CCSDT solution were analyzed for the (symmetric) bond stretches in hydrogen fluoride and water. In summary, the fourth-order CCSD(T--4) model was found to offer a stable, albeit relatively costly alternative to, e.g., the CCSD(T) model, returning the smallest errors among all of the tested models. In the present study, we will extend the analysis in Ref. \citenum{triples_pert_theory_jcp_2015}, not only by conducting a similar assessment for a test set of 18 atoms and small organic radicals, but also by reporting results for higher-order (fifth- and sixth-order) models of the CCSD(T--$n$) series, for total (equilibrium) electronic energies and for the symmetric bond stretches in the closed-shell water molecule and the open-shell amidogen radical, as examples of systems at the onset of bond-cleavage for which static correlation effects become increasingly important. In the second part of the present series~\cite{open_shell_quadruples_arxiv_2016}, we conduct a related study for the CCSDT(Q--$n$) quadruples series~\cite{ccsd_pert_theory_jcp_2014,quadruples_pert_theory_jcp_2015}, which is defined---theoretically on par with the CCSD(T--$n$) series---as an order expansion of a bivariational CCSDTQ energy Lagrangian around a CCSDT zeroth-order expansion point.\\

For closed-shell systems, given that the ground state is dominated by a single determinant, the solution from a restricted HF (RHF) calculation is adequate as the reference for a subsequent treatment of correlation effects. However, for ground states with $S \neq 0$, or even for molecules which are most appropriately described by (biradical) resonance structures, the traditional approach has been to invoke different orbitals for different spins, which is what is done in a more general unrestricted HF (UHF) calculation~\cite{pople_nesbet_uhf_jcp_1954}. However, a UHF determinant will in general not correspond to a pure spin state, i.e., the resulting wave function will usually not be an eigenfunction of the $\hat{S}^{2}$ operator. Such contamination of the spin, arising from spurious and highly unphysical contributions from other spin states, is commonly considered a major drawback of any UHF-based method, although the inclusion of correlation effects will generally help to reduce the contamination~\cite{jensen_uhf_cpl_1990}. For this reason, a restricted open-shell HF (ROHF) trial function is often preferred over the corresponding UHF one whenever the UHF expectation value $\langle \hat{S}^{2} \rangle$ deviates significantly from the correct value, i.e., whenever $\langle \hat{S}^{2} \rangle - S_z(S_z + 1) \neq 0$~\cite{szalay_uhf_rohf_cc_jcp_2004}. On the other hand, the lack of spin polarization in ROHF solutions may occasionally manifest itself as a drawback and render UHF the better choice. For this reason, the choice between the two types of open-shell references is an intricate one, and the assumption that ROHF-based correlation methods should always be preferred over corresponding UHF-based methods thus cannot be expected to hold for all radicals.

For both parts of the present series, we have deliberately chosen to limit our focus to correlated treatments on top of the UHF and ROHF references, as these are the two trial functions which are most commonly and routinely used for open-shell systems. For the sake of completeness, however, we acknowledge that the use of orbitals from Brueckner~\cite{chiles_dykstra_brueckner_cc_jcp_1981} or general orbital-optimized~\cite{scuseria_schaefer_oo_cc_cpl_1987} CC has previously been shown to reduce spin contamination in some cases~\cite{krylov_spin_cont_jcp_2000,taube_bartlett_jcp_1_2008} as well as to alleviate possible symmetry-breaking problems~\cite{stanton_gauss_brueckner_cc_jcp_1992}. In addition, the use of Kohn-Sham orbitals has been advocated as an inexpensive alternative to ordinary UHF references~\cite{byrd_head_gordon_open_shell_cc_jpca_2001}. However, an investigation of the possible gain from using any of these non-standard determinants falls outside the scope of the present work.

%
%
\section{Computational details}\label{com_details_section}

All of the perturbative CC triples models tested here give corrections to the CCSD model, as formulated for either a UHF or an ROHF reference~\cite{rittby_bartlett_rohf_ccsd_jpc_1988}. As in Ref. \citenum{triples_pert_theory_jcp_2015}, the natural reference target energy, against which we wish to compare the various models, will be that of the CCSDT model~\cite{ccsdt_paper_1_jcp_1987,*ccsdt_paper_2_cpl_1988,watts_bartlett_rohf_ccsdt_jcp_1990}. This choice of reference is motivated not only by the fact that the CCSDT model is the theoretical yardstick for all of the approximative triples models---for restricted as well as unrestricted trial functions---but also the fact that any spin contamination of the wave function is practically removed at the CCSDT level of theory. In Table \ref{spin_cont_table}, CC/cc-pVTZ expectation values of $\langle \hat{S}^{2} \rangle$ are summarized for all the members of our open-shell test set (due to the frozen-core approximation, the results in Table \ref{spin_cont_table} do not converge onto the theoretical values in the limit of full correlation among the valence electrons)~\cite{stanton_s2_exp_values_jcp_1994}. As may be recognized from the UHF- and ROHF-based results, the CCH and CN radicals (both doublets) as well as O$_2$ (triplet) differ from the rest of the test set, which we will further discuss in Sections \ref{ccsdpt_subsection} and \ref{ccsd_tn_subsection}.

\begin{table}[H]
      \caption{Orbital-unrelaxed CC/cc-pVTZ expectation values of $\langle \hat{S}^{2} \rangle$ for UHF- and ROHF-based CCSD, CCSDT, and CCSDTQ, reported relative to the theoretical values ($\Delta = S_z(S_z + 1) - \langle \hat{S}^{2} \rangle$). All units are $\Delta / 10^{-2}$.}
\label{spin_cont_table}
\small
{\begin{tabular}{l||r|r|r|r||r|r|r}
\hline\hline
\multicolumn{1}{c||}{Molecule} & \multicolumn{4}{c||}{UHF} & \multicolumn{3}{c}{ROHF} \\
\cline{2-8}
& \multicolumn{1}{c|}{$\Delta_{\text{SCF}}$} & \multicolumn{1}{c|}{$\Delta_{\text{CCSD}}$} & \multicolumn{1}{c|}{$\Delta_{\text{CCSDT}}$} & \multicolumn{1}{c||}{$\Delta_{\text{CCSDTQ}}$} & \multicolumn{1}{c|}{$\Delta_{\text{CCSD}}$} & \multicolumn{1}{c|}{$\Delta_{\text{CCSDT}}$} & \multicolumn{1}{c}{$\Delta_{\text{CCSDTQ}}$} \\
\hline
C	&	-0.9037	&	-0.0140	&	-0.0014	&	-0.0012	&	-0.0107	&	-0.0002	&	-0.0001	\\
CCH	&	-37.5055	&	-0.8609	&	-0.0677	&	-0.0031	&	-0.2335	&	-0.0065	&	-0.0004	\\
CF	&	-1.1917	&	-0.0353	&	-0.0003	&	-0.0002	&	-0.0276	&	0.0007	&	0.0004	\\
CH	&	-0.8894	&	-0.0170	&	-0.0014	&	-0.0006	&	-0.0111	&	-0.0005	&	0.0000	\\
CH$_{2}$ ($^3\text{B}_1$)	&	-1.6226	&	-0.0341	&	-0.0015	&	-0.0008	&	-0.0225	&	-0.0003	&	0.0000	\\
CH$_3$	&	-0.8245	&	-0.0220	&	-0.0007	&	-0.0003	&	-0.0137	&	-0.0001	&	0.0000	\\
CN	&	-38.9106	&	-0.4086	&	-0.0312	&	-0.0022	&	-0.1383	&	-0.0016	&	-0.0005	\\
F	&	-0.3414	&	-0.0054	&	-0.0005	&	-0.0004	&	-0.0027	&	-0.0001	&	0.0000	\\
HCO	&	-1.5251	&	-0.0519	&	0.0018	&	-0.0001	&	-0.0438	&	0.0023	&	0.0001	\\
HO$_2$\textsuperscript{\emph{a}}	&	-0.9080	&	-0.0281	&	-0.0016	&	-0.0005	&	-0.0236	&	-0.0009	&	-0.0003	\\
N	&	-0.6090	&	-0.0137	&	-0.0017	&	-0.0015	&	-0.0102	&	-0.0002	&	0.0000	\\
NH	&	-1.5265	&	-0.0304	&	-0.0017	&	-0.0010	&	-0.0187	&	-0.0004	&	0.0000	\\
NH$_2$	&	-0.8857	&	-0.0206	&	-0.0010	&	-0.0005	&	-0.0120	&	-0.0002	&	0.0000	\\
NO	&	-3.9060	&	-0.0613	&	-0.0026	&	-0.0011	&	-0.0362	&	-0.0014	&	-0.0005	\\
O	&	-0.7200	&	-0.0126	&	-0.0011	&	-0.0009	&	-0.0071	&	-0.0002	&	0.0000	\\
O$_2$	&	-4.2460	&	-0.0705	&	-0.0188	&	-0.0025	&	-0.0660	&	-0.0168	&	-0.0016	\\
OF	&	-1.8253	&	-0.0566	&	0.0010	&	-0.0005	&	-0.0380	&	0.0019	&	0.0000	\\
OH	&	-0.6027	&	-0.0126	&	-0.0007	&	-0.0004	&	-0.0070	&	-0.0002	&	0.0000	\\
\hline\hline
  \end{tabular}}
    \\
\textsuperscript{\emph{a}} CC/cc-pVDZ expectation values.
\end{table}
Besides the CCSD(T--2), CCSD(T--3), and CCSD(T--4) models of Ref. \citenum{triples_pert_theory_jcp_2015}, we will also assess the accuracy of the fifth- and sixth-order models of the CCSD(T--$n$) triples series---the CCSD(T--5) and CCSD(T--6) models, respectively---for which closed- and open-shell results will be presented for the first time. Furthermore, the results for all three references (RHF, UHF, and ROHF) will be compared to corresponding results for the CCSD(T) model, the accuracy of which will be initially evaluated. In terms of computational cost, the CCSD(T) and CCSD(T--2) models both scale with system size, $N$, as $\mathcal{O}(N^7)$ in the non-iterative step, while all of the higher-order models of the CCSD(T--$n$) series ($n>2$) scale as $\mathcal{O}(N^8)$ with an increasing number of cost-determining contractions upon moving up through the series, ultimately making the difference between a perturbative and a full CCSDT calculation negligible for orders $n>6$.

On par with both of our previous analyses for perturbative CC triples and quadruples expansions~\cite{triples_pert_theory_jcp_2015,quadruples_pert_theory_jcp_2015}, which were carried out for a test set of 17 small closed-shell molecules at their equilibrium geometries~\bibnote{Closed-shell test set: H$_2$O; H$_2$O$_2$; CO; CO$_2$; C$_2$H$_2$; C$_2$H$_4$; CH$_2$ ($^{1}\text{A}_{1}$); CH$_2$O; N$_2$; NH$_3$; N$_2$H$_2$; HCN; HOF; HNO; F$_2$; HF; O$_3$. Geometries are listed in Ref. \citenum{mest}.}, we will here perform a statistical analysis for a test set of 18 atoms and small radicals, all optimized at the all-electron CCSD(T)/cc-pVQZ level of theory and listed in Table \ref{spin_cont_table}. We note that this test set has previously been used in the calibration of the HEAT thermochemical model~\cite{heat_1_jcp_2004,*heat_2_jcp_2006,*heat_3_jcp_2008}. For measuring the numerical performance of each of the models, we report results in terms of (i) the relative recovery of the contribution to the CCSDT correlation energy from triple excitations (i.e., the CCSDT--CCSD correlation energy difference) as well as (ii) the actual deviation from this difference. All of the RHF-based calculations have been performed within the {\textsc{ncc}} module~\cite{ncc} of the {\textsc{cfour}} quantum chemical program package~\cite{cfour}, while the recently developed {\textsc{aquarius}} program~\cite{aquarius} has been used for all of the UHF-/ROHF-based calculations (in both the ROHF-based CCSD(T--$n$) and CCSD(T) calculations, a semicanonical reference is used)~\cite{watts_rohf_ccsdpt_jcp_1993}.~\bibnote{For all the models, the procedure was always first to transform to a semicanonical representation and then freezing the core in a subsequent step, cf. Ref. \citenum{watts_rohf_ccsdpt_jcp_1993}.}~\bibnote{A note on the CCSD(T--$n$) implementations for an ROHF reference; by using semicanonical orbitals, the only change to the existing UHF-based implementation is the presence of off-diagonal Fock matrix elements ($f_{ia}$ and $f_{ai}$). As these elements are counted as first-order in MBPT, we include them in the similarity-transformed fluctuation potential. Since the transformed fluctuation potential already contains matrix elements of the same type, arising from contractions between the bare fluctuation potential and the cluster operators, no additional changes to the existing code are necessary.} The correlation-consistent cc-pVTZ basis set~\cite{dunning_1_orig,*dunning_5_core} is used throughout for all of the reported valence-electron (frozen-core) results.

%
%
\section{Results}\label{results_section}

In this section, we start by investigating the performance of the CCSD(T) model against the CCSDT model for UHF and ROHF references in Section \ref{ccsdpt_subsection}, while corresponding results for the CCSD(T--$n$) models are reported in Section \ref{ccsd_tn_subsection}. In both cases, we report statistical error measures generated from the individual results, cf. the supplementary material~\bibnote{See supplementary material at [AIP URL] for individual recoveries and deviations.}, and these are, in turn, compared to RHF-based results for the closed-shell test set of Refs. \citenum{triples_pert_theory_jcp_2015} and \citenum{quadruples_pert_theory_jcp_2015}. Finally, limited-range CCSD(T--$n$) and CCSD(T) potential energy surfaces for the symmetric vibrational modes in the water molecule and amidogen radical are reported in Section \ref{pes_subsection}.

%
%
\subsection{The CCSD(T) model}\label{ccsdpt_subsection}
\begin{figure}
        \centering
        \begin{subfigure}[b]{0.47\textwidth}
                \includegraphics[width=\textwidth,bb=0 2 447 362]{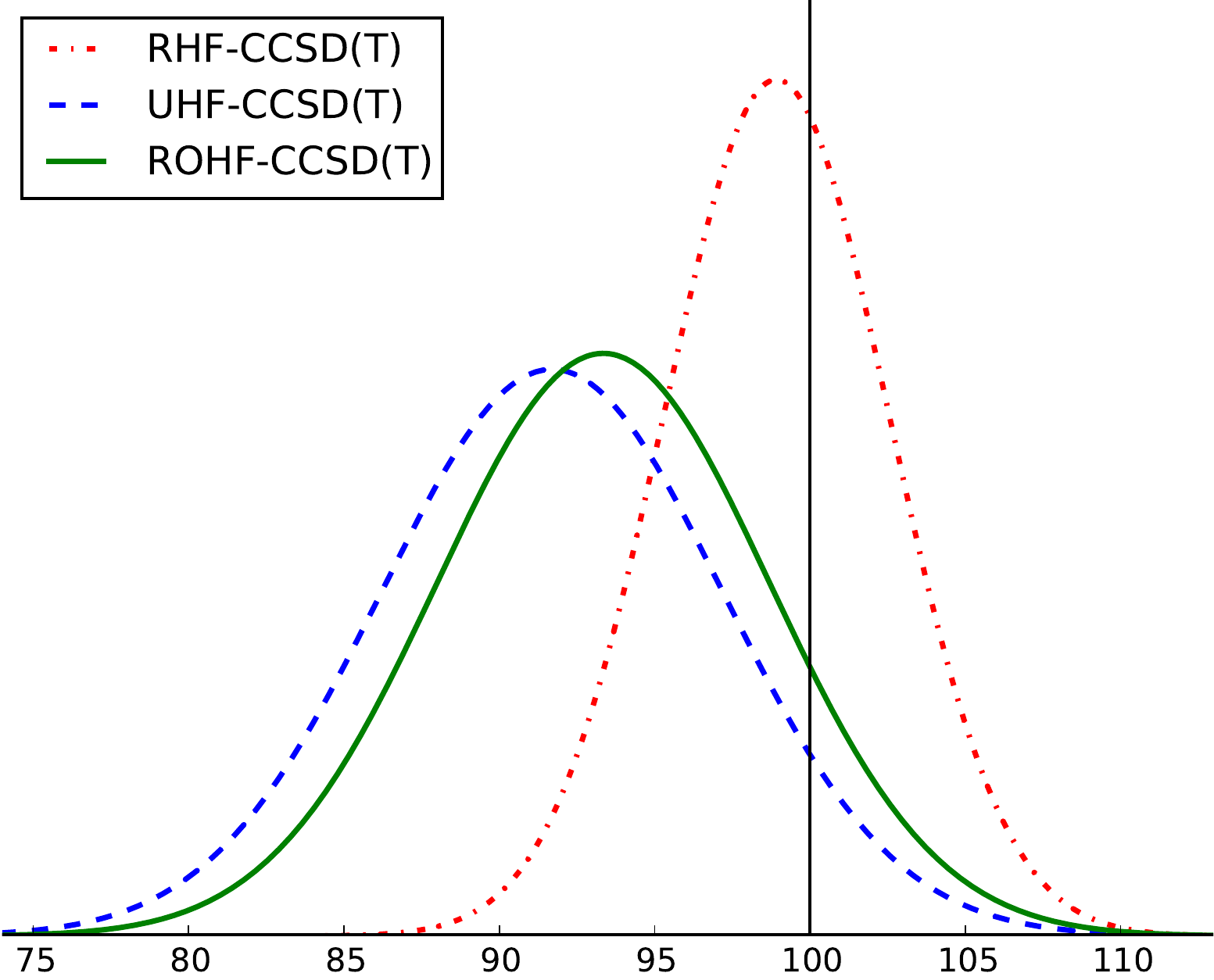}
                \caption{Recoveries}
                \label{ccsdpt_recoveries_figure}
        \end{subfigure}%
        ~ 
        \begin{subfigure}[b]{0.47\textwidth}
                \includegraphics[width=\textwidth,bb=0 2 447 362]{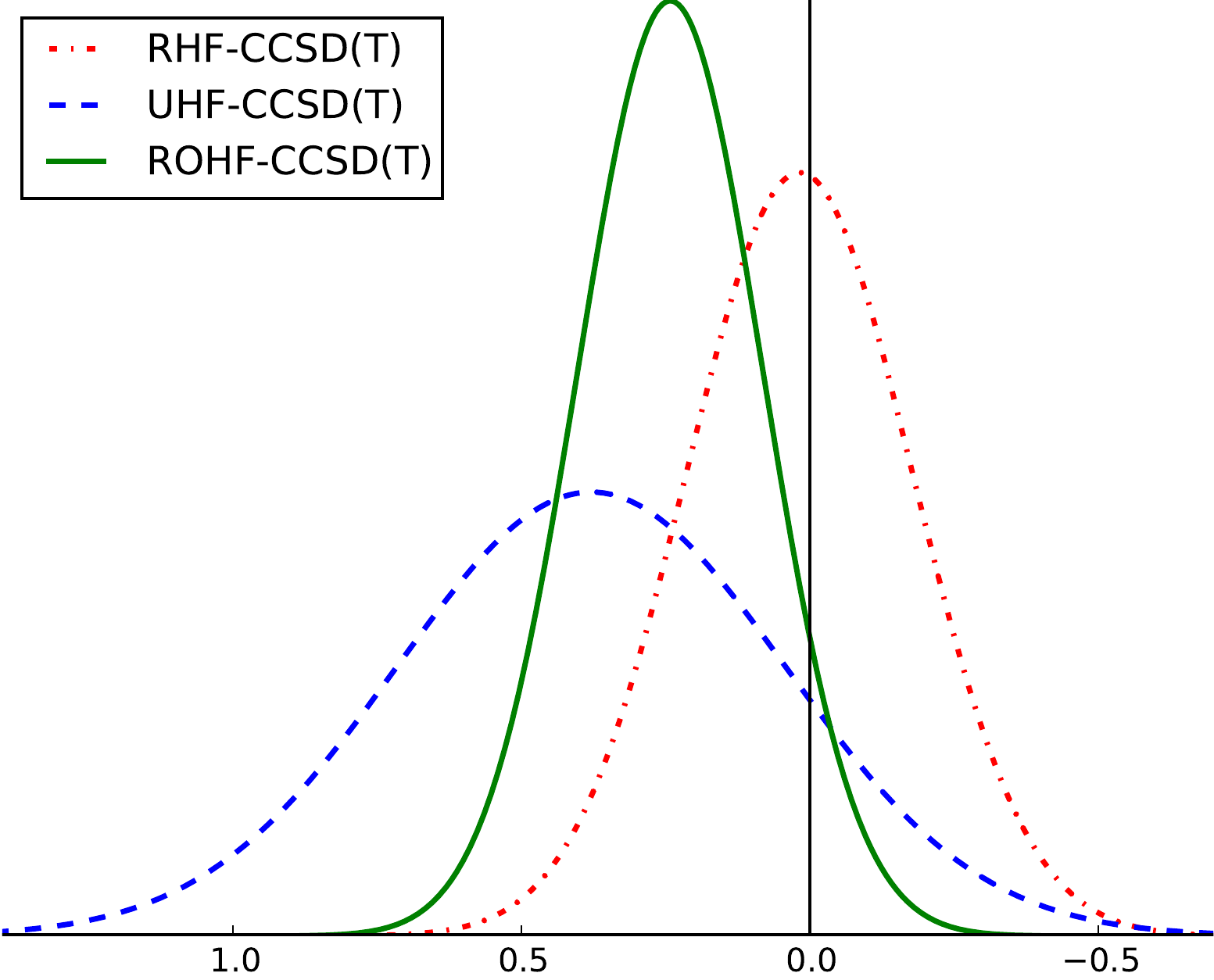}
                \caption{Deviations}
                \label{ccsdpt_abs_diff_figure}
        \end{subfigure}
        \caption{Normal distributions of the recoveries of (in percent (\%), Figure \ref{ccsdpt_recoveries_figure}) and deviations from (in kcal/mol, Figure \ref{ccsdpt_abs_diff_figure}) CCSDT--CCSD frozen-core/cc-pVTZ correlation energy differences for RHF, UHF, and ROHF references.}
        \label{ccsdpt_figure}
\end{figure}
In Figure \ref{ccsdpt_figure}, we present CCSD(T) mean recoveries of/deviations from the CCSDT triples contribution to the correlation energy as well as standard deviations around the means for the RHF, UHF, and ROHF calculations, all compactly depicted in terms of normal distributions. Inspecting first the results in Figure \ref{ccsdpt_recoveries_figure}, two things are immediately clear: first, the consistency of the CCSD(T) model in recovering CCSDT triples effects for closed-shell systems is clearly worsened in the transition from closed- to open-shell systems, and second, the statistical results for the two open-shell references appear very similar. Whereas the first observation agrees well with the aforementioned discrepancy between closed- and open-shell CCSD(T) results for equilibrium geometries and harmonic frequencies, the second is slightly surprising, taking into account that some of the UHF reference determinants may be significantly spin contaminated (cf. Table \ref{spin_cont_table}). This becomes clearer upon inspecting the actual deviations in Figure \ref{ccsdpt_abs_diff_figure}. Here, a pronounced difference is observed between the UHF- and ROHF-based CCSD(T) results where the latter, i.e., the agreement of the CCSD(T) and CCSDT models for ROHF determinants, correspond more closely to the RHF-based results (RHF-CCSD(T) versus RHF-CCSDT for the comparable closed-shell test set). By comparing the individual entries of Tables S5 and S6 of the supplementary material, in which results are presented for the deviations from UHF- and ROHF-CCSDT correlation energies, respectively, we find that the large difference between the UHF- and ROHF-based results in Figure \ref{ccsdpt_abs_diff_figure} is primarily due to the inclusion of two radicals into the test set which are heavily plagued by spin contamination, namely the challenging isoelectronic cases of the ethynyl and cyano radicals, cf. Table \ref{spin_cont_table}. 
\begin{figure}
        \centering
        \begin{subfigure}[b]{0.47\textwidth}
                \includegraphics[width=\textwidth,bb=0 2 447 362]{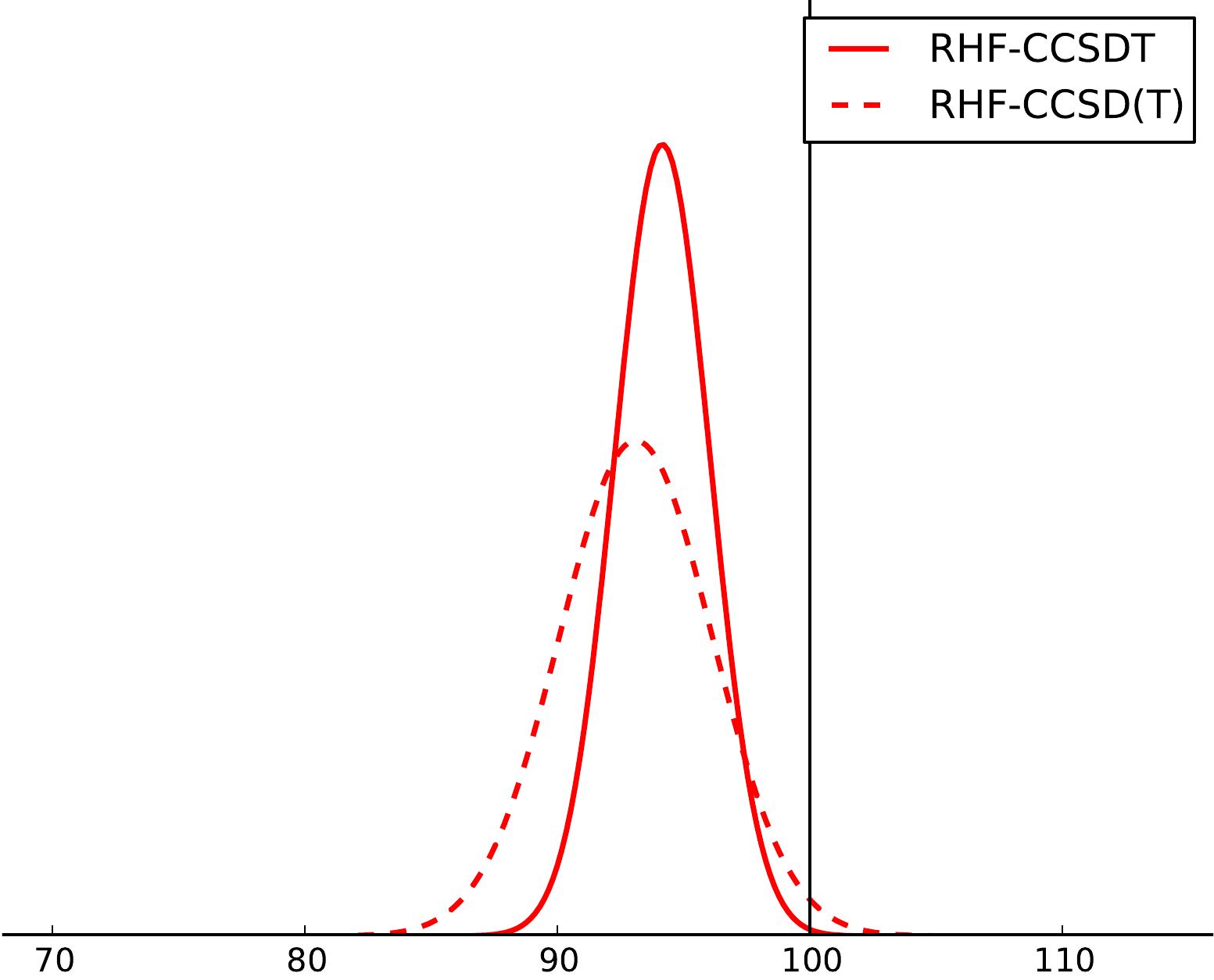}
                \caption{RHF recoveries}
                \label{ccsdtq_recoveries_rhf_figure}
        \end{subfigure}%
        ~ 
        \begin{subfigure}[b]{0.47\textwidth}
                \includegraphics[width=\textwidth,bb=0 2 447 362]{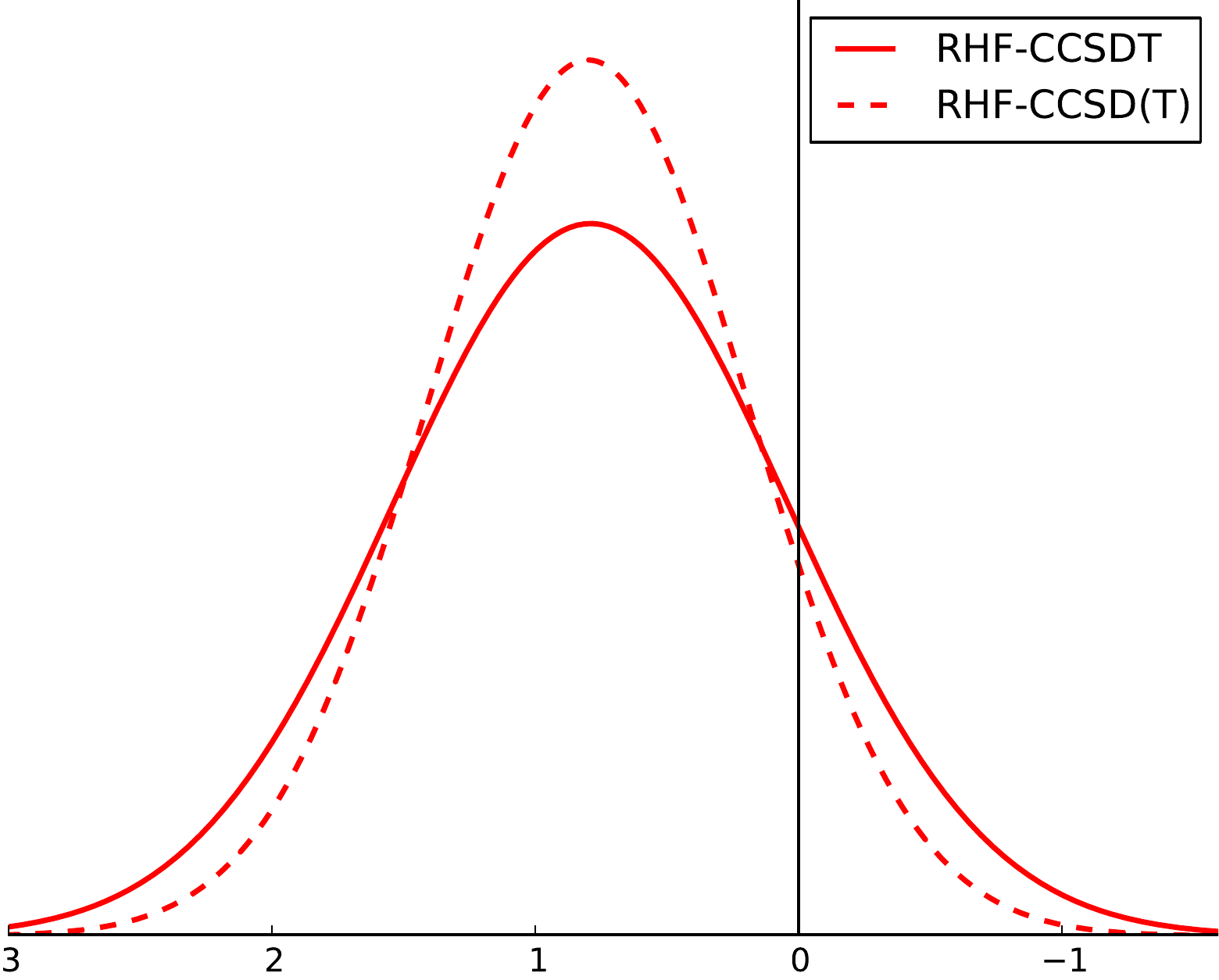}
                \caption{RHF deviations}
                \label{ccsdtq_abs_diff_rhf_figure}
        \end{subfigure}
        \begin{subfigure}[b]{0.47\textwidth}
                \includegraphics[width=\textwidth,bb=0 2 447 362]{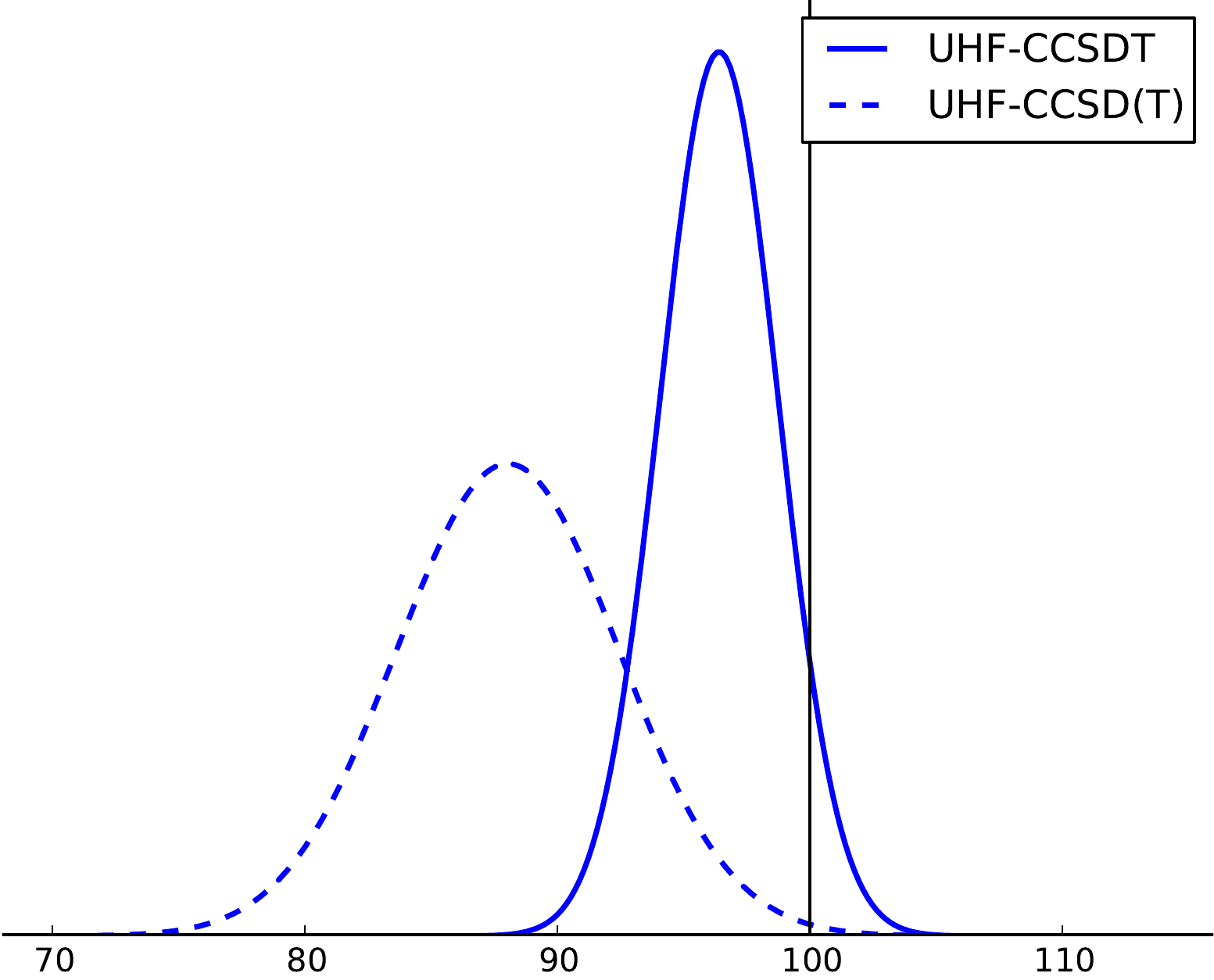}
                \caption{UHF recoveries}
                \label{ccsdtq_recoveries_uhf_figure}
        \end{subfigure}%
        ~ 
        \begin{subfigure}[b]{0.47\textwidth}
                \includegraphics[width=\textwidth,bb=0 2 447 362]{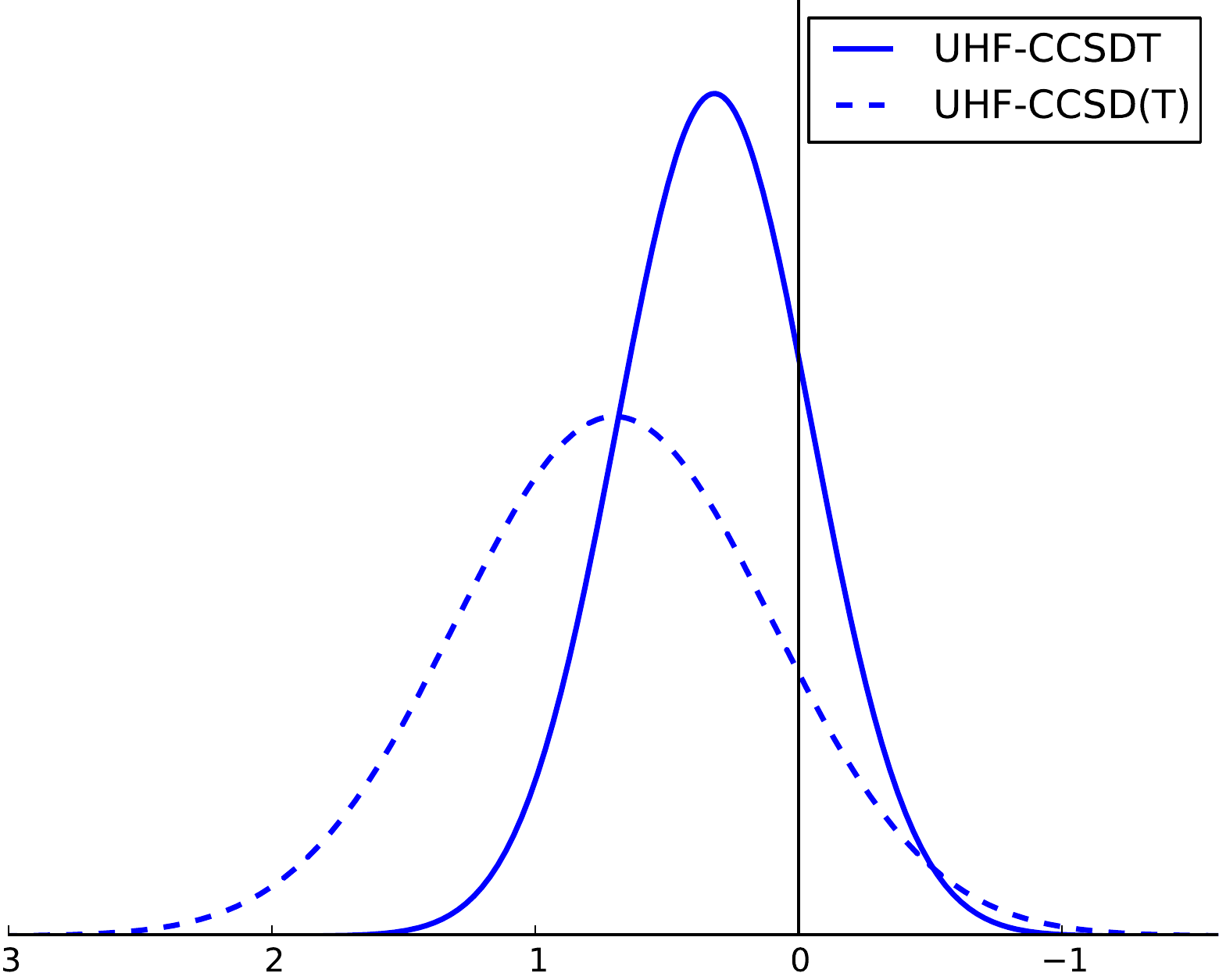}
                \caption{UHF deviations}
                \label{ccsdtq_abs_diff_uhf_figure}
        \end{subfigure}
        \begin{subfigure}[b]{0.47\textwidth}
                \includegraphics[width=\textwidth,bb=0 2 447 362]{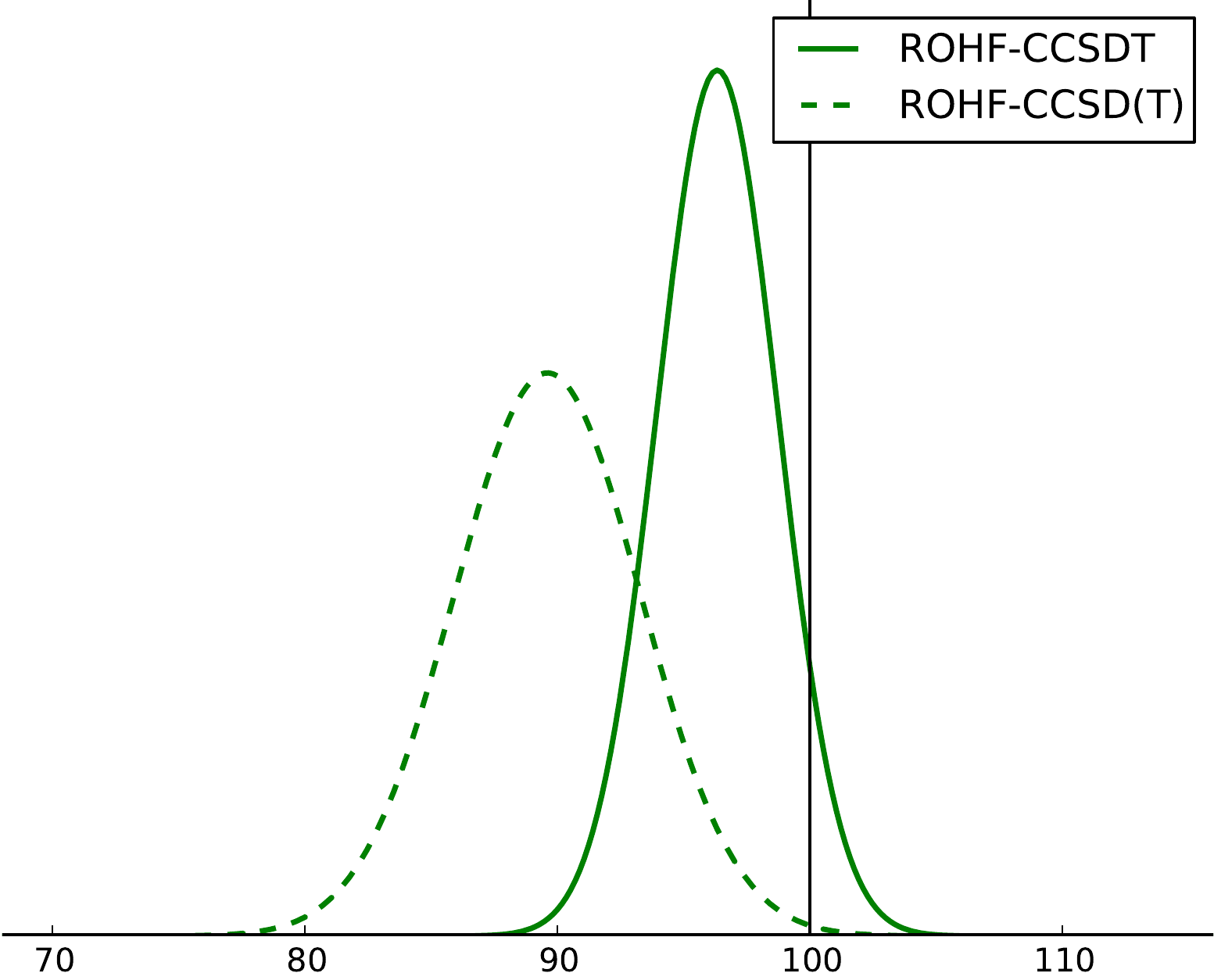}
                \caption{ROHF recoveries}
                \label{ccsdtq_recoveries_rohf_figure}
        \end{subfigure}%
        ~ 
        \begin{subfigure}[b]{0.47\textwidth}
                \includegraphics[width=\textwidth,bb=0 2 447 362]{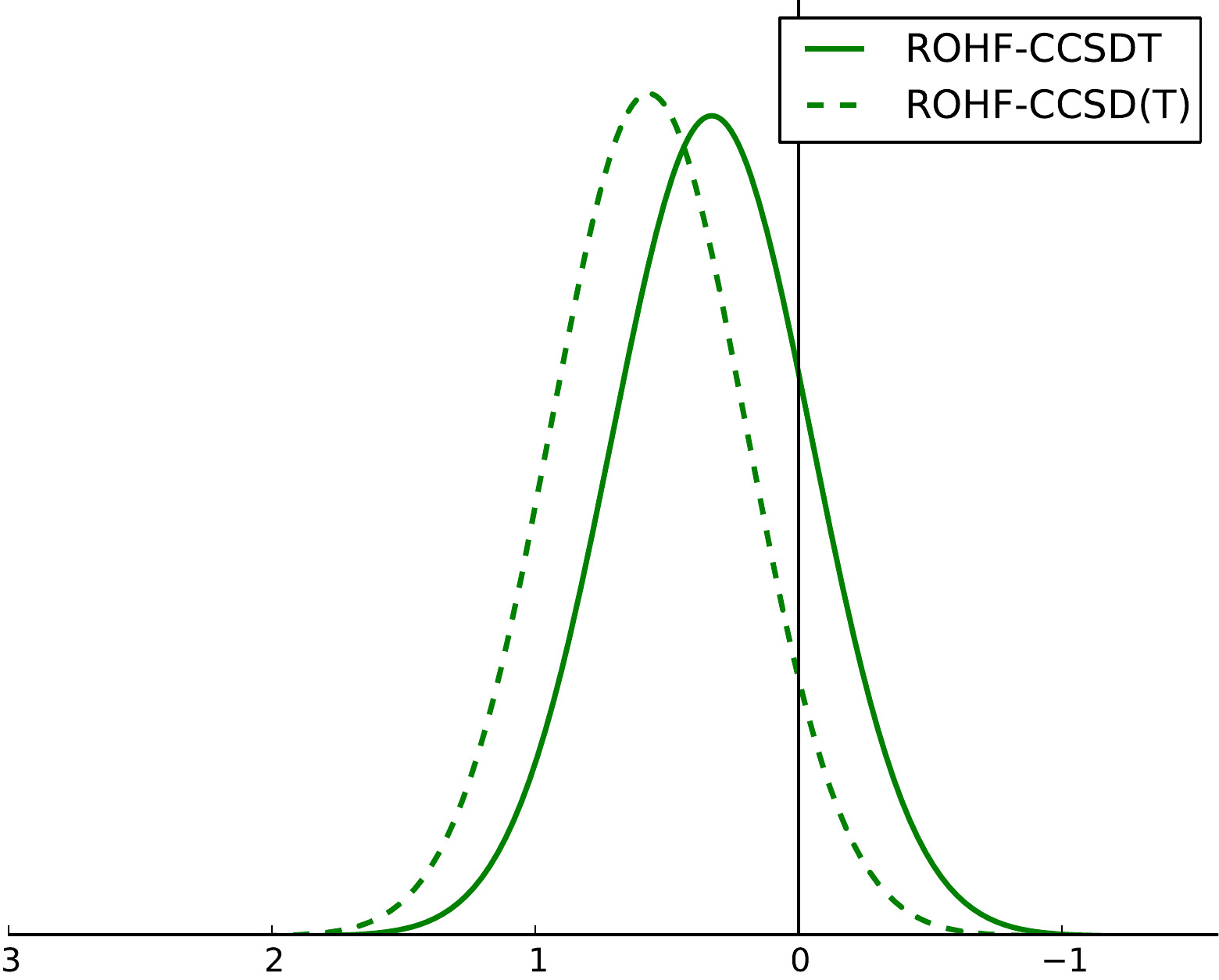}
                \caption{ROHF deviations}
                \label{ccsdtq_abs_diff_rohf_figure}
        \end{subfigure}
        \caption{Normal distributions of the recoveries of (in percent (\%), Figures \ref{ccsdtq_recoveries_rhf_figure}, \ref{ccsdtq_recoveries_uhf_figure}, and \ref{ccsdtq_recoveries_rohf_figure}) and deviations from (in kcal/mol, Figures \ref{ccsdtq_abs_diff_rhf_figure}, \ref{ccsdtq_abs_diff_uhf_figure}, and \ref{ccsdtq_abs_diff_rohf_figure}) CCSDTQ--CCSD frozen-core/cc-pVTZ correlation energy differences for RHF, UHF, and ROHF references.}
        \label{ccsdtq_figure}
\end{figure}
In fact, CCH and CN behave radically different from the rest of the set at the UHF-CCSD(T) level, as has already been pointed out in Ref. \citenum{heat_1_jcp_2004}, and they are the only two radicals for which the ROHF-CCSDT triples contribution is larger than the corresponding UHF-CCSDT contribution (although less than 1.0 kJ/mol). If one were to exclude these two outliers from the statistical analysis (UHF-CCSD(T) deviations of 0.90 and 1.44 kcal/mol from the CCSDT--CCSD differences for CCH and CN, respectively), the UHF and ROHF distributions in Figure \ref{ccsdpt_abs_diff_figure} would essentially overlap, while it would have no positive effect on the results for the recoveries in Figure \ref{ccsdpt_recoveries_figure}.

As mentioned in Section \ref{intro_section}, the remarkable performance of the CCSD(T) model in chemical applications for closed-shell molecules (in particular for energy differences and properties) is often attributed to an overestimation of effects from triple excitations to such an extent that RHF-CCSD(T) results are in better agreement with more advanced CC models, e.g., the CCSDTQ model or even the FCI limit, than with its natural reference, the CCSDT model. By comparing individual RHF-based CCSD(T) and CCSDT correlation energies to corresponding RHF-CCSDTQ results in Tables S7 and S8 of the supplementary information, this is indeed recognized as an occasional, although far from general feature of the model in the context of total energies (unlike for molecular properties). This is graphically illustrated in Figure \ref{ccsdtq_figure}, which gives CCSD(T) and CCSDT mean recoveries of/deviations from the combined CCSDTQ triples and quadruples contribution to the correlation energy as well as standard deviations around the means. For the closed-shell test set, the molecules that overestimate the CCSDT triples contribution the most (N$_2$, O$_3$, and CO$_2$) only manage to ``recover'' $18.0 \%$, $12.7 \%$, and $12.6 \%$, respectively, of the total contribution to the CCSDTQ energy from quadruple excitations. On the other hand, we note how this potential asset of the model is in general absent for open-shell calculations, in comparing UHF and ROHF-based CCSD(T) and CCSDT results to corresponding UHF- and ROHF-CCSDTQ results (cf. Figure \ref{ccsdtq_figure} and Tables S7 and S8 of the supplementary information). In fact, only for one open-shell species (the oxygen molecule, and only in a triple-$\zeta$ basis) does the CCSD(T) model give a marginal overestimation of the CCSDT triples contribution.

%
%
\subsection{The CCSD(T--$n$) series}\label{ccsd_tn_subsection}
\begin{figure}
        \centering
        \begin{subfigure}[b]{0.47\textwidth}
                \includegraphics[width=\textwidth,bb=0 0 488 378]{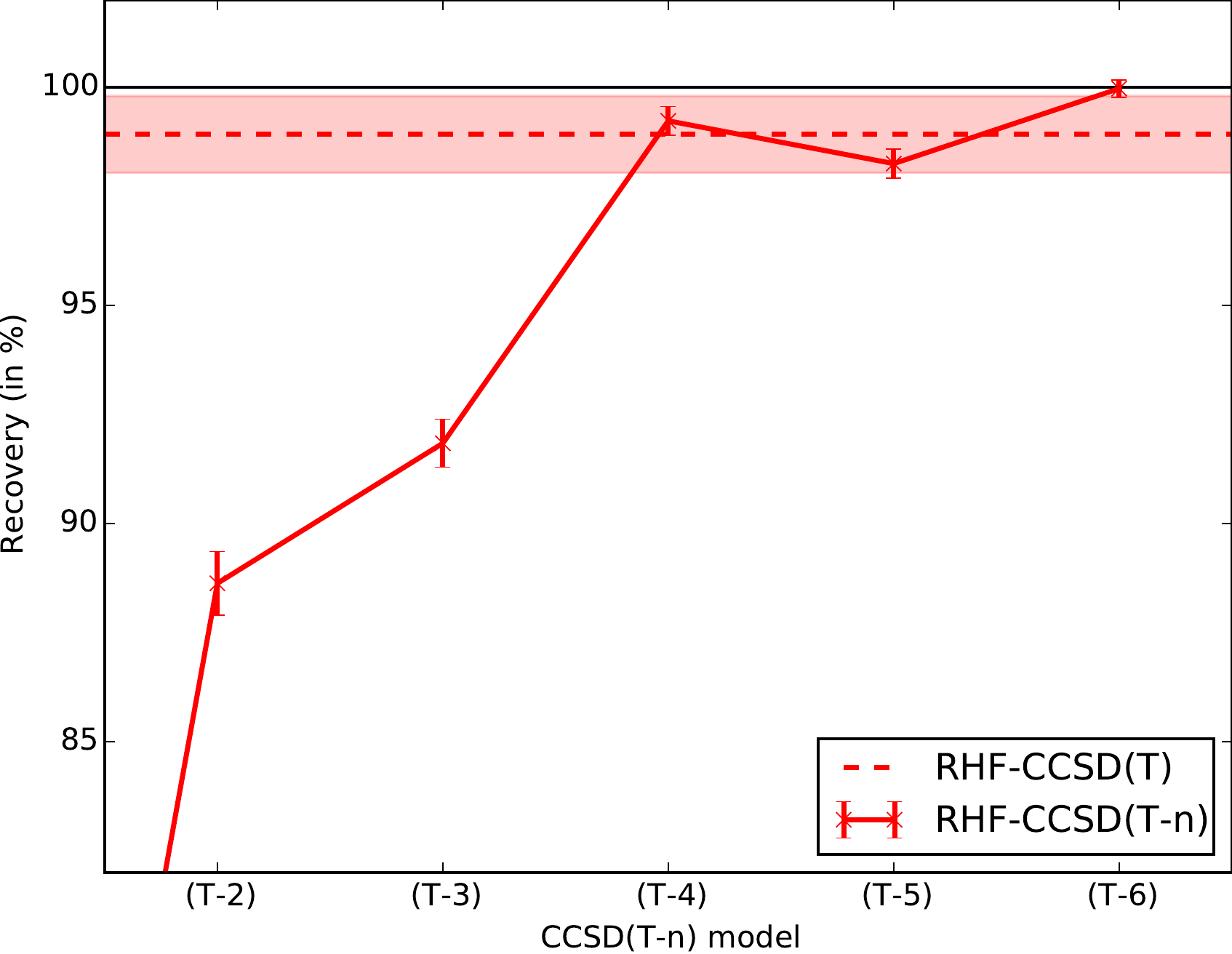}
                \caption{RHF recoveries}
                \label{t_n_recoveries_rhf_figure}
        \end{subfigure}%
        \hspace{0.4cm} 
        \begin{subfigure}[b]{0.47\textwidth}
                \includegraphics[width=\textwidth,bb=0 0 488 378]{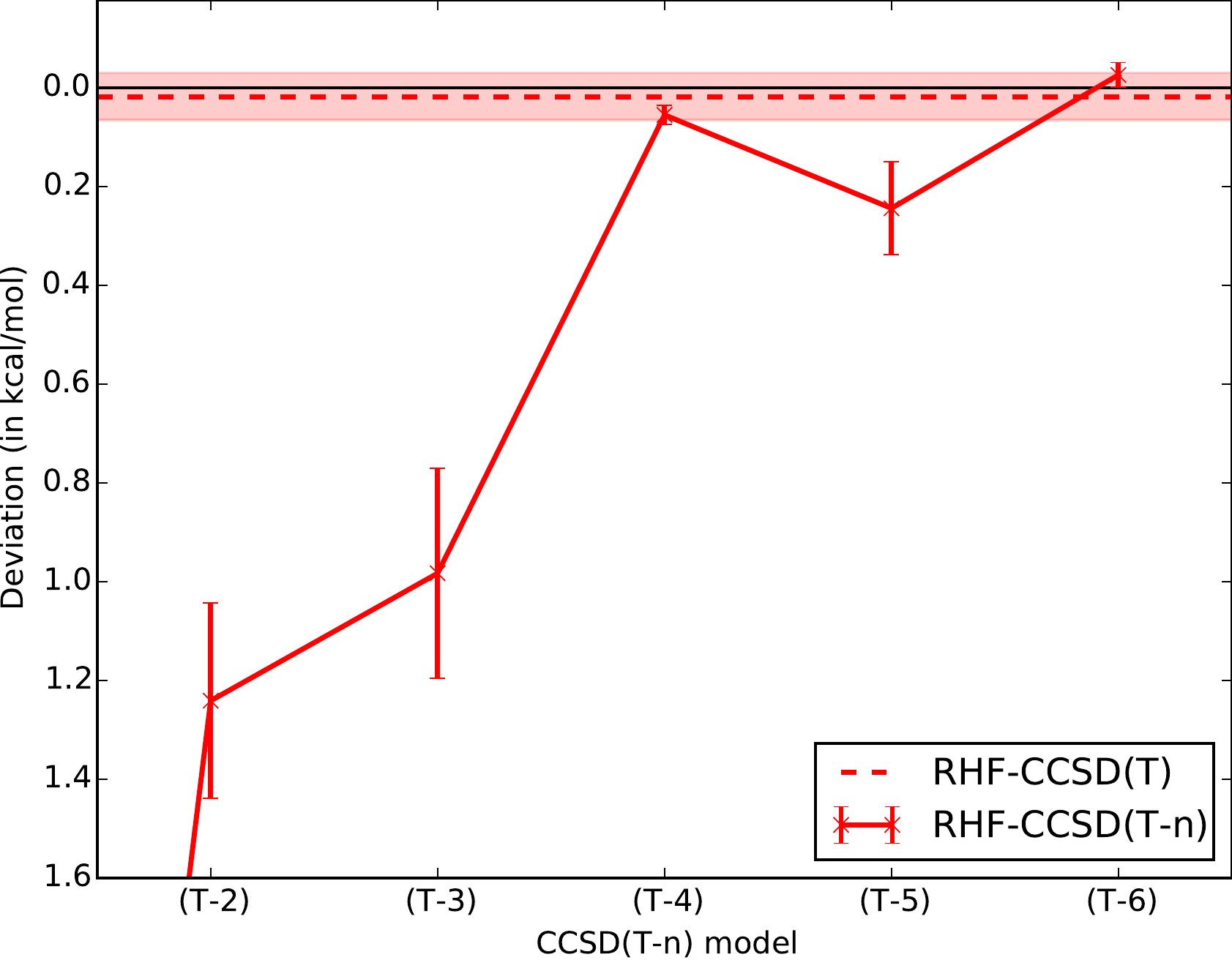}
                \caption{RHF deviations}
                \label{t_n_abs_diff_rhf_figure}
        \end{subfigure}
        \begin{subfigure}[b]{0.47\textwidth}
                \includegraphics[width=\textwidth,bb=0 0 488 378]{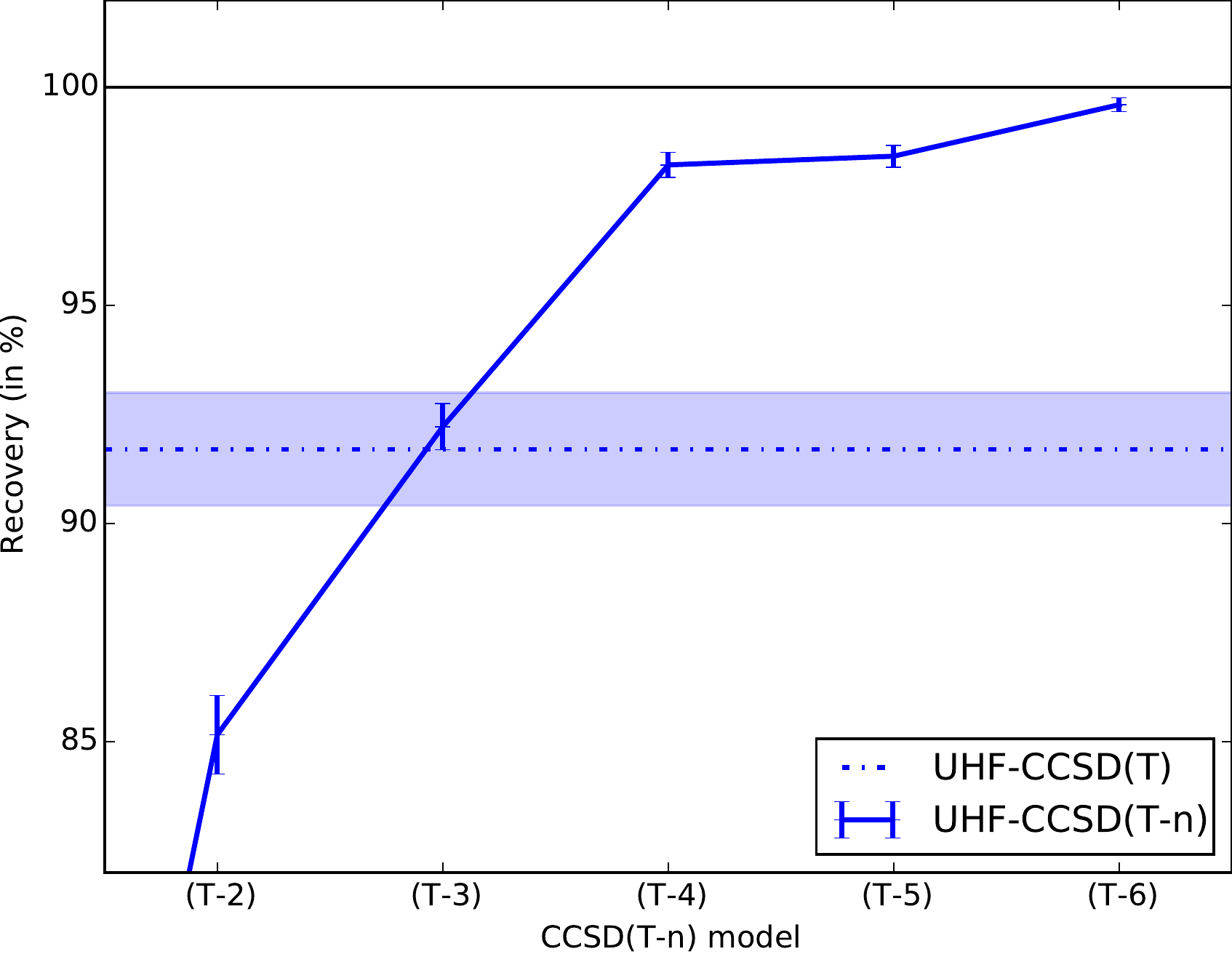}
                \caption{UHF recoveries}
                \label{t_n_recoveries_uhf_figure}
        \end{subfigure}%
        \hspace{0.4cm} 
        \begin{subfigure}[b]{0.47\textwidth}
                \includegraphics[width=\textwidth,bb=0 0 488 378]{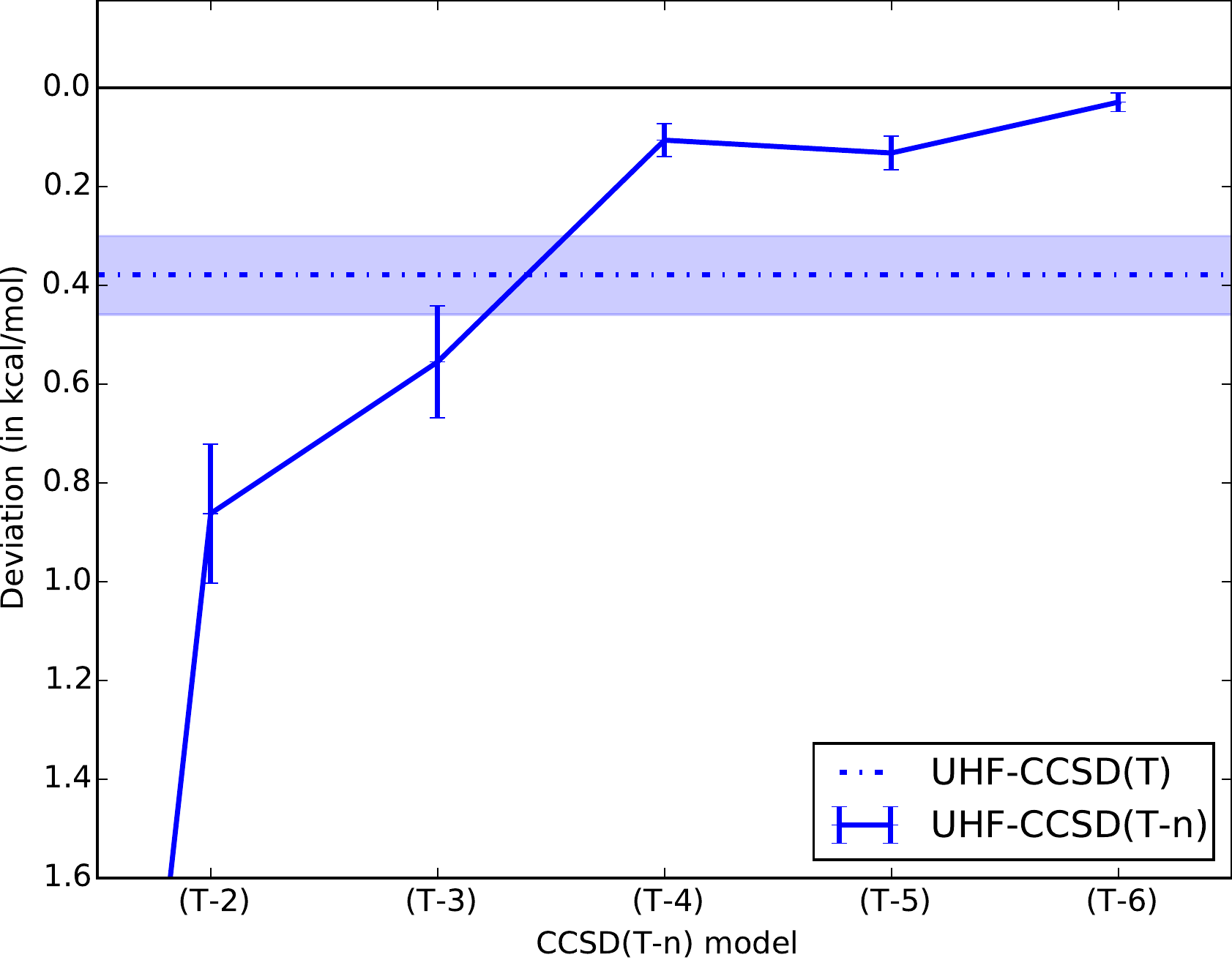}
                \caption{UHF deviations}
                \label{t_n_abs_diff_uhf_figure}
        \end{subfigure}
        \begin{subfigure}[b]{0.47\textwidth}
                \includegraphics[width=\textwidth,bb=0 0 488 378]{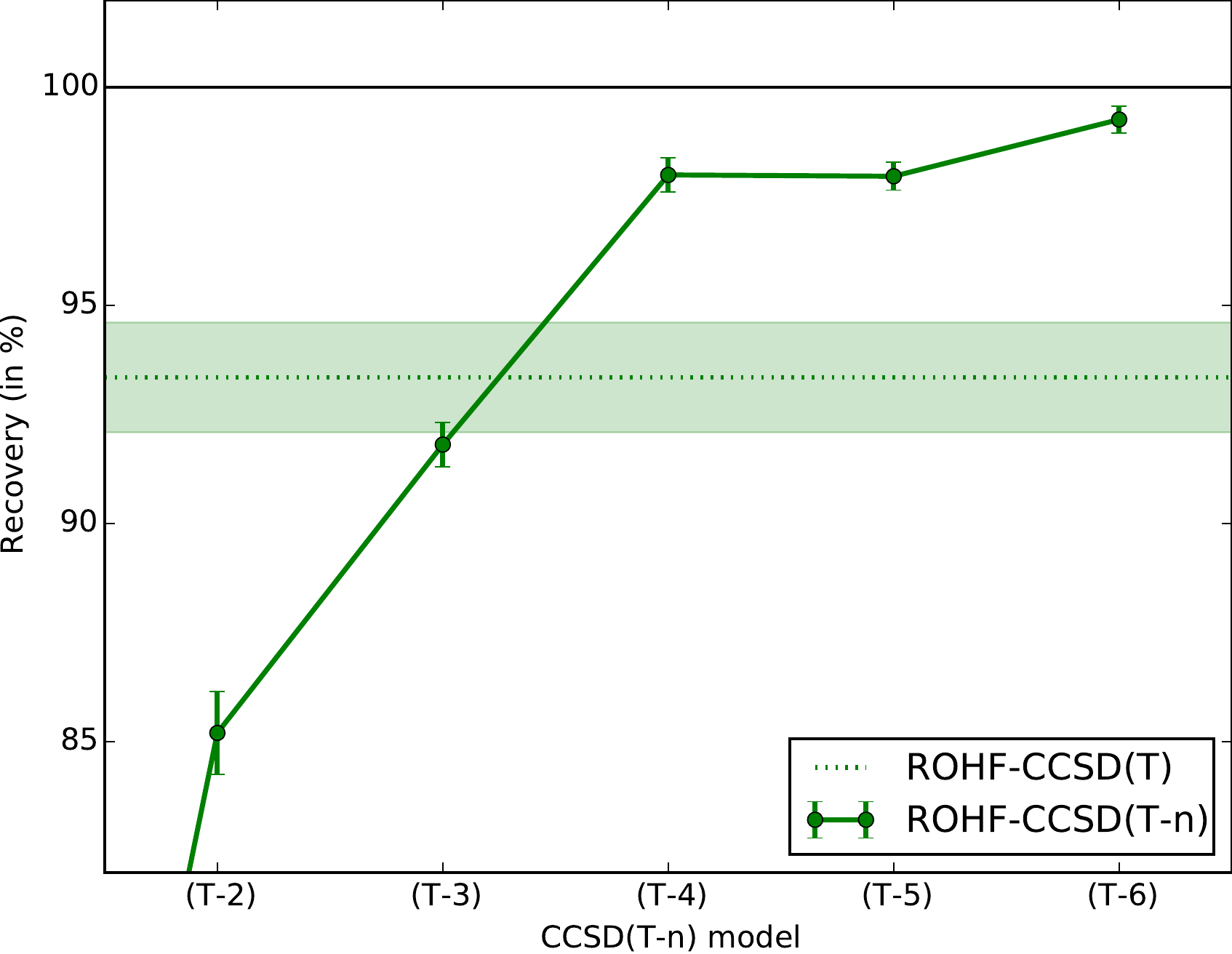}
                \caption{ROHF recoveries}
                \label{t_n_recoveries_rohf_figure}
        \end{subfigure}%
        \hspace{0.4cm} 
        \begin{subfigure}[b]{0.47\textwidth}
                \includegraphics[width=\textwidth,bb=0 0 488 378]{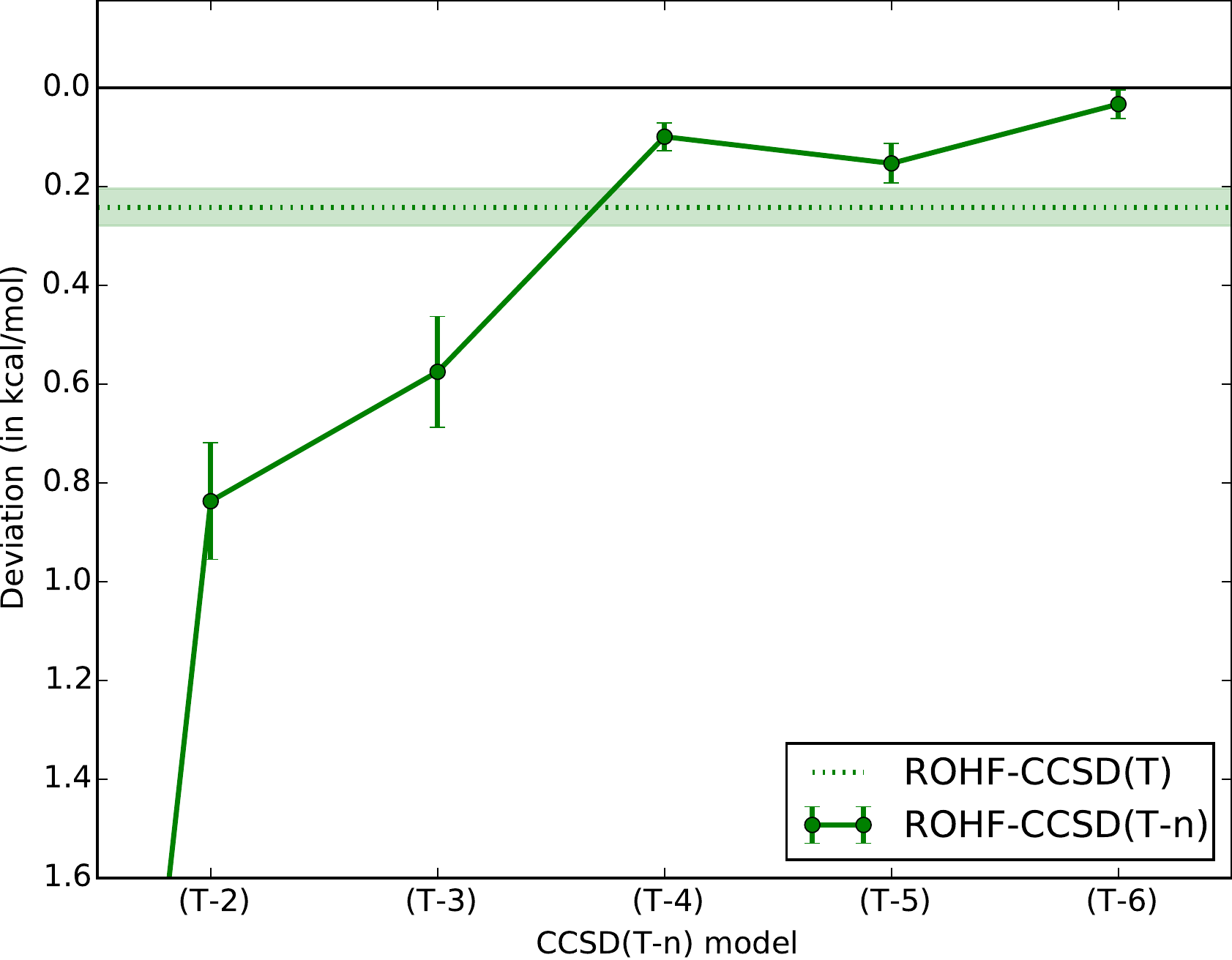}
                \caption{ROHF deviations}
                \label{t_n_abs_diff_rohf_figure}
        \end{subfigure}
        \caption{Recoveries of (in percent (\%), Figures \ref{t_n_recoveries_rhf_figure}, \ref{t_n_recoveries_uhf_figure}, and \ref{t_n_recoveries_rohf_figure}) and deviations from (in kcal/mol, Figures \ref{t_n_abs_diff_rhf_figure}, \ref{t_n_abs_diff_uhf_figure}, and \ref{t_n_abs_diff_rohf_figure}) CCSDT--CCSD frozen-core/cc-pVTZ correlation energy differences for RHF, UHF, and ROHF references. The error bars show the standard error of the mean, and the CCSD(T) recoveries/deviations of Figure \ref{ccsdpt_figure} have been included for comparison (for this model, the standard error is depicted as a colored interval centred around the mean).}
        \label{t_n_figure}
\end{figure}

The results in Figure \ref{ccsdpt_figure}, in particular the rather poor performance of the CCSD(T) model for the two open-shell references, now raise an evident follow-up question: is it possible to improve upon the CCSD(T) model without having to resort to iterative triples models? For instance, if any perturbative model is to be an improvement over the CCSD(T) model for, say, simple atomization energies of CCSDT quality, it would have to challenge (if not improve) the level of accuracy that may be achieved for closed-shell species while improving upon it for open-shell species. In Ref. \citenum{triples_pert_theory_jcp_2015}, we compared the CCSD(T) model to the second-, third-, and fourth-order models of the recently proposed CCSD(T--$n$) perturbation series (alongside a broad range of alternative triples models) and found that only the fourth-order model (CCSD(T--4)) was capable of improving the accuracy---albeit at an increased computational cost---as measured against the target CCSDT triples model. For this reason, we will compare the UHF- and ROHF-CCSD(T) models to the second- through sixth-order models of the CCSD(T--$n$) series for both of these references. In Figure \ref{t_n_figure}, results are presented for all of these models, again with RHF-based results included for a full comparison (the CCSD(T) results for RHF, UHF, and ROHF references are the same of those in Figure \ref{ccsdpt_figure}). Inspecting first the recovery of the CCSDT triples contribution in Figures \ref{t_n_recoveries_rhf_figure}, \ref{t_n_recoveries_uhf_figure}, and \ref{t_n_recoveries_rohf_figure}, we note how the theoretically predicted convergence of the CCSD(T--$n$) series towards the CCSDT target energy is numerically confirmed through sixth order in the perturbation, and, most importantly, how this holds true not only for closed-shell species, but regardless of the spin of the ground state. The three different curves for RHF, UHF, and ROHF are remarkably similar, with the most notable differences being the RHF results for the CCSD(T--2) and CCSD(T--4) models, which are most likely slightly better than what should be expected, as may be seen by comparing these to the corresponding open-shell results. Furthermore, we recognize---as was hinted to in Ref. \citenum{triples_pert_theory_jcp_2015}---how the ordinary staircase convergence of HF-based perturbation theory (MBPT) is preserved in the CCSD(T--$n$) series, which is based on a CCSD zeroth-order reference; while the even-ordered corrections to the CCSD energy are large, the odd-ordered corrections are generally much smaller in magnitude, although the fifth-order corrections often work to rectify the somewhat disproportionate, yet rather consistent results achieved at fourth order. 

As far as the deviations of the CCSD(T--$n$) models from the CCSDT triples contribution in Figures \ref{t_n_abs_diff_rhf_figure}, \ref{t_n_abs_diff_uhf_figure}, and \ref{t_n_abs_diff_rohf_figure} are concerned, the picture is observed to be much the same. We note how the RHF-based CCSD(T--2) and CCSD(T--3) results are worse than the corresponding UHF- and ROHF-based ones, both in terms of mean and standard deviations, but how this discrepancy between the closed- and open-shell descriptions vanishes at fourth order. We find that this is due in particular to the ozone molecule, cf. Table S4 of the supplementary material, which, at the CCSD(T--2) and CCSD(T--3) levels, has a contribution from triples excitations that deviates by more than 4.0 kcal/mol from the CCSDT reference value. At the CCSD(T--4) level, however, this deviation is reduced to $0.1$ kcal/mol as a result of the relaxation and higher-order triples effects which enter the CCSD(T--$n$) series at fourth order~\cite{triples_pert_theory_jcp_2015}. Finally, we note from Table S6 of the supplementary material that O$_2$, when described by an ROHF reference determinant, is an odd case for the CCSD(T--$n$) models, giving a deviation at the CCSD(T--6)/cc-pVTZ level that is about six times larger than the second largest deviation. One possible reason for this is the relatively large deviation from the theoretical $\langle \hat{S}^{2} \rangle$ value left at the ROHF-CCSDT level for this molecule (cf. Table \ref{spin_cont_table}). As mentioned in Section \ref{ccsdpt_subsection}, the oxygen molecule is the only member of the open-shell test set for which the CCSD(T) model overestimates the CCSDT contribution from triple excitations, and the large differences exhibited by the models of the CCSD(T--$n$) series, which otherwise account satisfactorily for triples effects, thus strongly indicate that the ROHF-CCSD(T) result for this molecule might be rather artificial. 

%
%
\subsection{Calculations at distorted geometries}\label{pes_subsection}

In the course of a chemical reaction, a molecule is often distorted away from its equilibrium geometry along a given internal coordinate. In such situations, the stretching of bond(s) will introduce multiconfigurational character, that is, static correlation, into the wave function, which makes a mean-field HF description of the reference state increasingly worse. An immediate consequence of this is that the overall description of the system becomes exceedingly more difficult. Furthermore, the weights of higher-level excited determinants in the wave function will grow, making it inherently more demanding to apply perturbative models for the recovery of these pronounced effects. Thus, in order to further test the models of the CCSD(T--$n$) series---despite the fact that neither of these have been designed specifically to recover static correlation effects---we will here evaluate the models for the symmetric bond stretches in the closed-shell water (H$_2$O) molecule and the open-shell amidogen (NH$_2$) radical and compare them in terms of their non-parallelity errors with respect to the target CCSDT solution (the validity of which we will initially assess).

\begin{figure}
        \centering
        \begin{subfigure}[b]{0.47\textwidth}
                \includegraphics[width=\textwidth,bb=0 1 485 386]{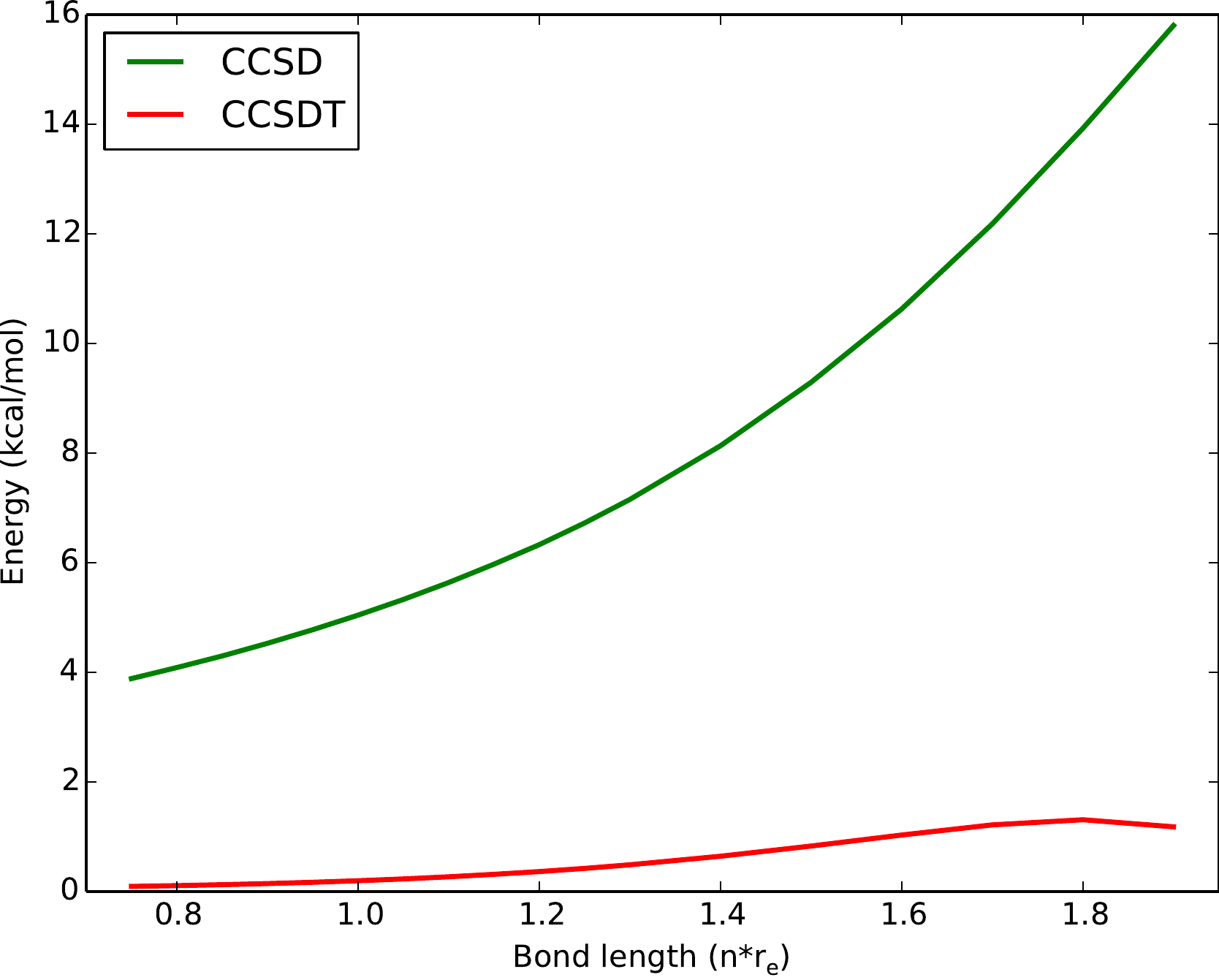}
                \caption{Deviation from CCSDTQ}
                \label{pes_h2o_ccsdtq_figure}
        \end{subfigure}%
        ~ 
        \begin{subfigure}[b]{0.47\textwidth}
                \includegraphics[width=\textwidth,bb=0 1 485 386]{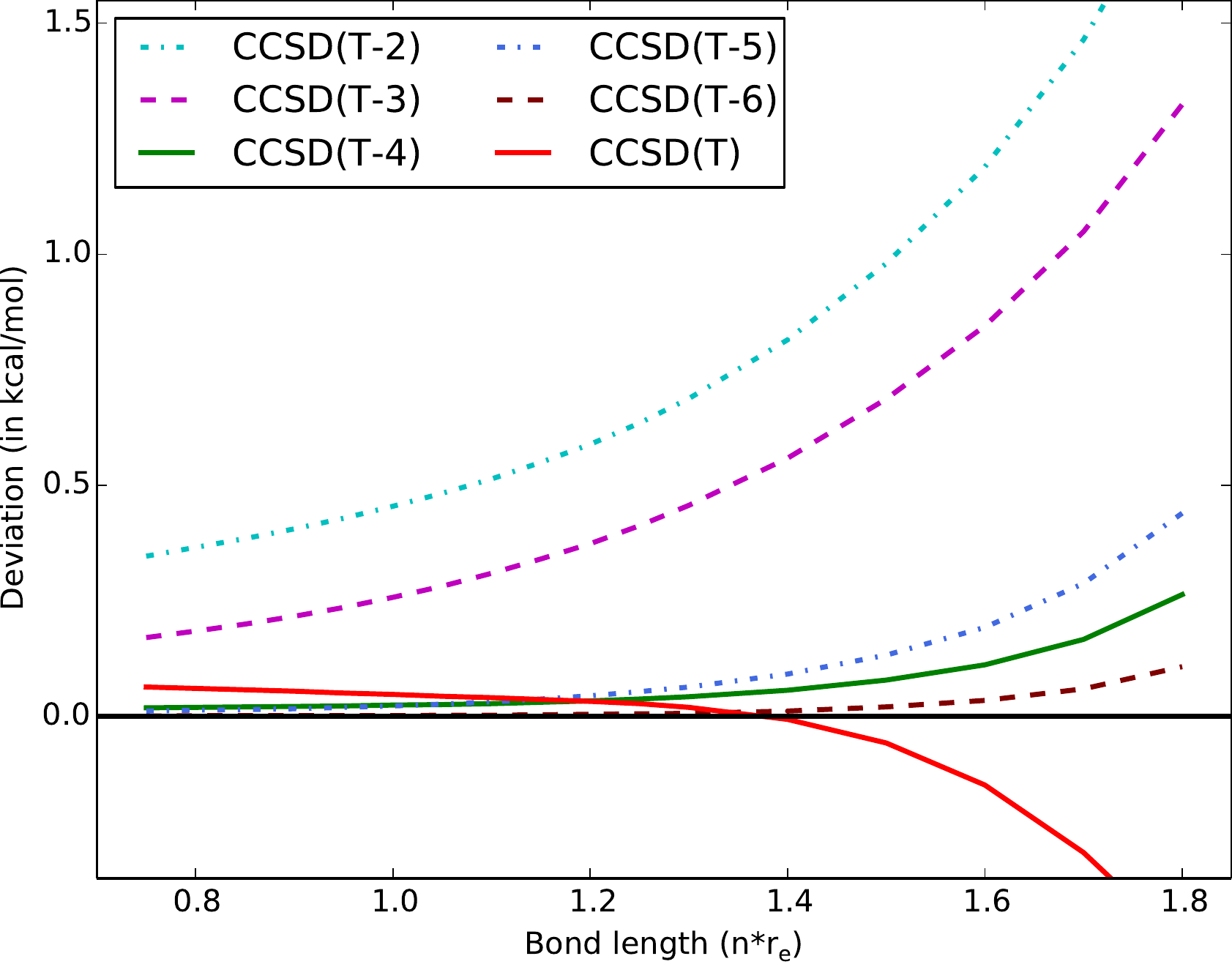}
                \caption{Deviation from CCSDT}
                \label{pes_h2o_ccsdt_figure}
        \end{subfigure}
        \caption{Deviation (in kcal/mol) from CCSDTQ and CCSDT frozen-core/cc-pVTZ correlation energies (Figures \ref{pes_h2o_ccsdtq_figure} and \ref{pes_h2o_ccsdt_figure}, respectively) for the water molecule at various bond lengths. The equilibrium bond length ($\text{r}_{\text{e}}$) is $95.7$ pm.}
        \label{pes_h2o_figure}
\end{figure}
\begin{figure}
        \centering
        \begin{subfigure}[b]{0.47\textwidth}
                \includegraphics[width=\textwidth,bb=0 1 485 386]{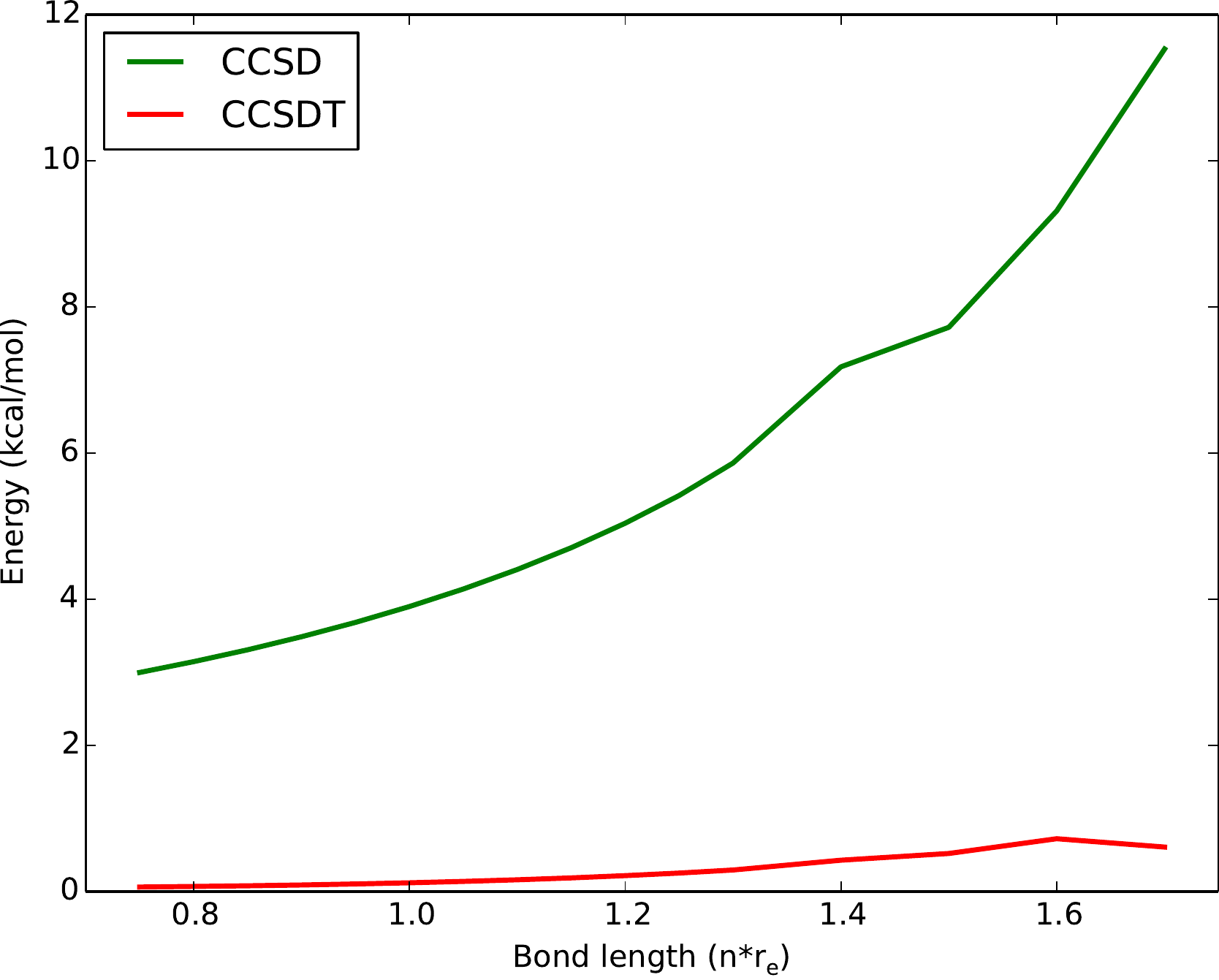}
                \caption{Deviation from CCSDTQ}
                \label{pes_nh2_ccsdtq_figure}
        \end{subfigure}%
        ~ 
        \begin{subfigure}[b]{0.47\textwidth}
                \includegraphics[width=\textwidth,bb=0 1 485 386]{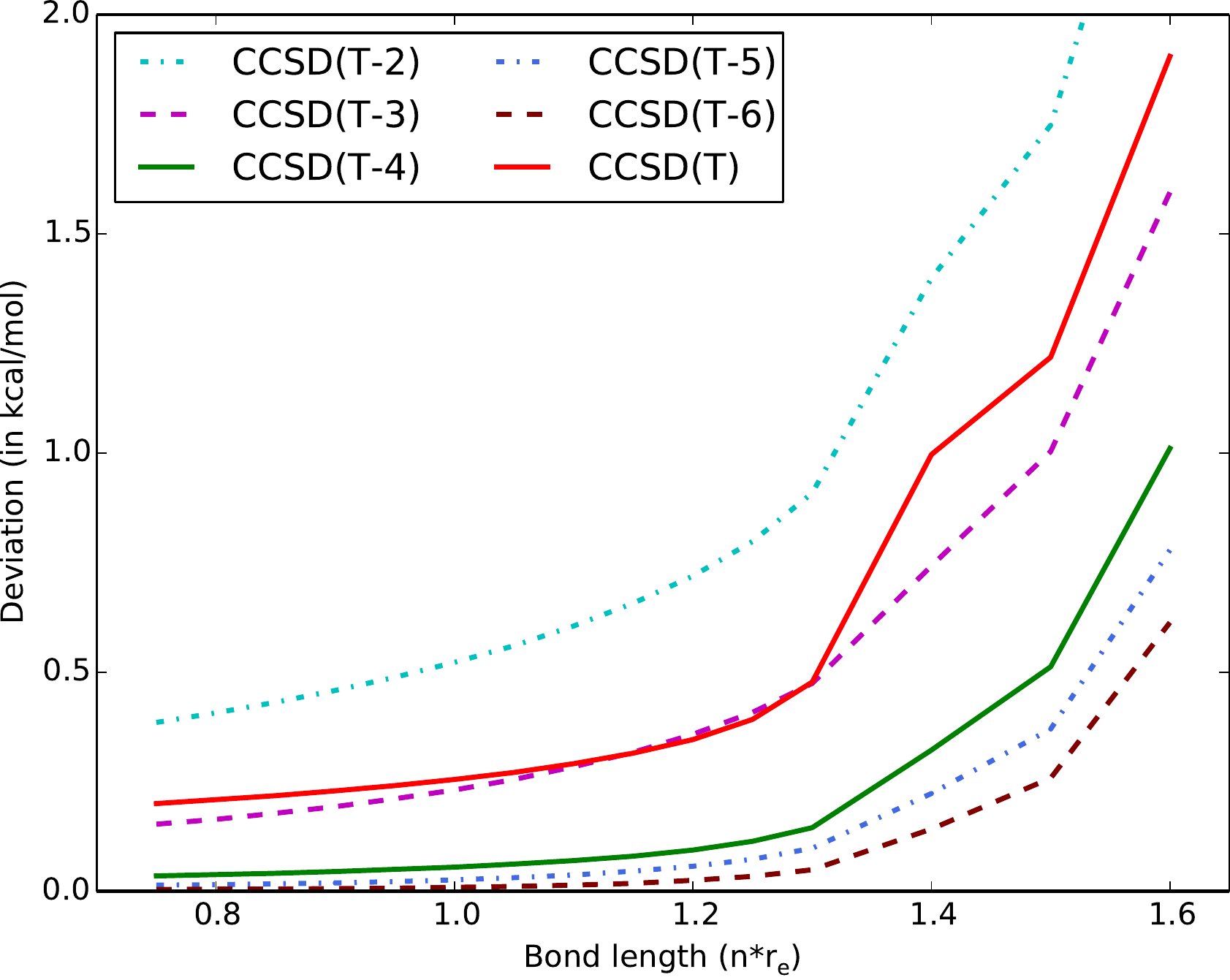}
                \caption{Deviation from CCSDT}
                \label{pes_nh2_ccsdt_figure}
        \end{subfigure}
        \caption{Deviation (in kcal/mol) from CCSDTQ and CCSDT frozen-core/cc-pVTZ correlation energies (Figures \ref{pes_nh2_ccsdtq_figure} and \ref{pes_nh2_ccsdt_figure}, respectively) for the amidogen radical at various bond lengths. The equilibrium bond length ($\text{r}_{\text{e}}$) is $102.3$ pm.}
        \label{pes_nh2_figure}
\end{figure}

For both water (RHF reference) and amidogen (UHF reference) in a cc-pVTZ basis ($C_{\text{2v}}$ point group), the error of the CCSDT solution with respect to that of the more advanced CCSDTQ model is positive and increases up to a certain bond length, at which the CCSDT solution collapses. Thus, it is therefore only of interest to test the performance of the CCSD(T--$n$) models within this region; for H$_2$O in Figure \ref{pes_h2o_ccsdtq_figure}, the CCSDT solution starts to collapse at an elongation of $80 \%$, while for NH$_2$ in Figure \ref{pes_nh2_ccsdtq_figure}, this critical point is met at an elongation of $60 \%$.

In Figures \ref{pes_h2o_ccsdt_figure} and \ref{pes_nh2_ccsdt_figure}, the total deviation from the correlation energy at the CCSDT/cc-pVTZ level of theory is reported for both of these stretches, again presented alongside corresponding results for the CCSD(T) model, which---like the models of the CCSD(T--$n$) series---is designed with the purpose of recovering dynamical correlation only. If we define a conservative confidence interval to be spanned by a maximum error of $\pm 0.2$ kcal/mol, we note how for the water stretch in Figure \ref{pes_h2o_ccsdt_figure}, the results for the CCSD(T--2) model all fall outside the confidence interval, the results for the CCSD(T--3) model nearly so, while the CCSD(T) results stay within up until a stretch of about $60\%$. Only the higher-order CCSD(T--$n$) models are capable of staying within the interval for all tested bond lengths with, e.g., the performance of the CCSD(T--4) model only showing signs of deteriorating near the very end of the tested interval of bond lengths. For the NH$_2$ stretch in Figure \ref{pes_nh2_ccsdt_figure}, both the native CCSD and CCSDT models as well as the CCSD(T--$n$) models are observed to behave similarly to the case of the H$_2$O stretch in Figure \ref{pes_h2o_figure}. However, while the CCSD and CCSDT models deviate less from the CCSDTQ solution than for H$_2$O, the CCSD(T--$n$) models (as measured against the CCSDT model) deviate slightly more. For the CCSD(T) model, on the other hand, the performance for the two test systems is markedly different; whereas the CCSD(T) results for NH$_2$ are similar to the corresponding CCSD(T--$n$) results, that is, the model consistently (and increasingly) underestimates the CCSDT triples effects, this is not the case for H$_2$O, for which the error is positive only up to a certain point (a stretch of about $40 \%$), at which the model suddenly starts to overestimate these effects.

Thus, whenever high-accuracy potential energy surfaces in the vicinity of the equilibrium geometry are required, as, e.g., when these form the basis for vibrational configuration interaction or coupled cluster response theory~\cite{christiansen_vcc_jcp_2004_1,*christiansen_vcc_jcp_2004_2,seidler_christiansen_vcc_resp_theory_jcp_2007} or in the determination of activation energies, the higher-order models of the CCSD(T--$n$) series might offer a stable non-iterative alternative to the CCSDT model. In particular, the fact that triples relaxation effects are accounted for in the fourth- and higher-orders models of the series will make these significantly more stable, albeit more expensive, than various other non-iterative models, as seen, e.g., from the comparison with the CCSD(T) model in Figures \ref{pes_h2o_figure} and \ref{pes_nh2_figure} and various other triples models in Ref. \citenum{triples_pert_theory_jcp_2015} (in which a similar analysis of the bond stretch in hydrogen fluoride is also presented).

%
%
\section{Summary and conclusion}\label{conclusion_section}

By evaluating frozen-core/cc-pVTZ total energies against CCSDT reference results, the performance of the second- through sixth-order models of the recently proposed CCSD(T--$n$) series of triples expansions for UHF and ROHF reference determinants has been statistically assessed for a test set of 18 atoms and small radicals. Furthermore, these results have been compared to {\bf{(i)}} the performance of the RHF-based variants of the models for a comparable test set of 17 closed-shell molecules as well as {\bf{(ii)}} corresponding CCSD(T) results for all three references. In summary, no differences are found between the treatments of closed- and open-shell species by the higher-order CCSD(T--$n$) models, neither for the size-intensive recovery of the CCSDT triples contribution nor the actual deviation from this. Moreover, the results presented here strongly indicate, as is the case for regular iterative CC models, that the higher-order CCSD(T--$n$) models are capable of significantly reducing the amount of spin contamination which may be present in underlying UHF trial functions, since no behavioural differences are observed between the UHF- and ROHF-based CCSD(T--$n$) results reported herein. These models, in particular the fourth-order CCSD(T--4) model, thus offer proper non-iterative, albeit more computationally expensive alternatives to the current {\it{de facto}} standard, the CCSD(T) model, for which the appraised numerical consistency in recovering CCSDT triples effects is worsened in the transition from closed- to open-shell systems, irrespective of the choice of reference determinant used for the latter type (UHF/ROHF).

In addition, the performance assessment of the CCSD(T--$n$) models has been extended to two prototypical bond stretches (the symmetric vibrational modes in the closed-shell H$_2$O molecule and open-shell NH$_2$ radical), as examples of systems for which static correlation effects become increasingly dominant. As for the calculations at equilibrium geometries, the higher-order models of the CCSD(T--$n$) triples series are found to offer results that are in significantly better agreement with the native CCSDT solution than the traditional CCSD(T) model for molecules at distorted geometries, as, e.g., encountered along reaction paths or at transition states. For instance, the performance of the CCSD(T--4) model is again found to be essentially independent of the reference used, and it will thus be capable of offering a balanced non-iterative treatment of closed- and open-shell species as required in, e.g., the determination of activation energies, even at intermediate bond lengths where alternative triples expansions may start to collapse.

%
%
\section*{Acknowledgments}

J. J. E. and P. J. acknowledge support from the European Research Council under the European Union's Seventh Framework Programme (FP/2007-2013)/ERC Grant Agreement No. 291371 and The Danish Council for Independent Research -- Natural Sciences. J. G. acknowledges financial support from the Deutsche Forschungsgemeinschaft (DFG GA 370/5-1). D. A. M. acknowledges support from the US National Science Foundation (NSF) under grant number ACI-1148125/1340293.

\newpage
\providecommand*\mcitethebibliography{\thebibliography}
\csname @ifundefined\endcsname{endmcitethebibliography}
  {\let\endmcitethebibliography\endthebibliography}{}


\begin{mcitethebibliography}{44}
\bibitem[Not()]{Note-3}
A note on the CCSD(T--$n$) implementations for an ROHF reference; by using
  semicanonical orbitals, the only change to the existing UHF-based
  implementation is the presence of off-diagonal Fock matrix elements ($f_{ia}$
  and $f_{ai}$). As these elements are counted as first-order in MBPT, we
  include them in the similarity-transformed fluctuation potential. Since the
  transformed fluctuation potential already contains matrix elements of the
  same type, arising from contractions between the bare fluctuation potential
  and the cluster operators, no additional changes to the existing code are
  necessary.\relax
\mciteBstWouldAddEndPunctfalse
\mciteSetBstMidEndSepPunct{\mcitedefaultmidpunct}
{}{\mcitedefaultseppunct}\relax
\EndOfBibitem
\end{mcitethebibliography}
\end{document}